\documentclass[fleqn,usenatbib]{mnras}

\usepackage{newtxtext,newtxmath}
\usepackage[T1]{fontenc}
\usepackage{longtable}
\DeclareRobustCommand{\VAN}[3]{#2}
\let\VANthebibliography\thebibliography
\def\thebibliography{\DeclareRobustCommand{\VAN}[3]{##3}\VANthebibliography}

\usepackage{graphicx}	% Including figure files
\usepackage{amsmath}	% Advanced maths commands 

\usepackage{enumitem}

\defcitealias{Poggianti2016}{P16}
\defcitealias{Yuan2020}{YH20}
\defcitealias{Vulcani2022}{V22}
\defcitealias{Yuan2022}{Y22}

\title[RPS and cluster dynamical states]{The effect of cluster dynamical state on ram-pressure stripping.}

\author[A. C. C. Louren\c{c}o et al.]{
Ana C. C. Louren\c{c}o,$^{1}$\thanks{E-mail: ana.lourenco@postgrado.uv.cl}
Y. L. Jaff\'e,$^{1}$
B. Vulcani,$^{2}$
A. Biviano,$^{3,4}$
B. Poggianti,$^{2}$
A. Moretti,$^{2}$
K. Kelkar,$^{1}$
\newauthor
J. P. Crossett,$^{1}$
M. Gitti,$^{5,6}$
R. Smith,$^{7}$
T. F. Lagan\'{a},$^{8}$
M. Gullieuszik,$^{2}$
A. Ignesti,$^{2}$
S. McGee,$^{9}$
\newauthor
A. Wolter,$^{10}$
S. Sonkamble,$^{2,12}$
and A. M\"uller$^{11}$
\\
$^{1}$Instituto de F\'isica y Astronomia, Universidad de Valpara\'iso, 1111 Gran Bretana, Valpara\'iso, Chile\\
$^{2}$INAF-Osservatorio astronomico di Padova, Vicolo Osservatorio 5, IT-35122 Padova, Italy\\
$^{3}$INAF-Osservatorio Astronomico di Trieste, via G. B. Tiepolo 11, 34143 Trieste, Italy\\
$^{4}$IFPU-Institute for Fundamental Physics of the Universe, via Beirut 2, 34014 Trieste, Italy\\
$^{5}$Dipartimento di Fisica e Astronomia, Università di Bologna, Via Gobetti 93/2, 40129 Bologna, Italy\\
$^{6}$INAF, Istituto di Radioastronomia di Bologna, Via Gobetti 101, 40129 Bologna, Italy\\
$^{7}$Departamento de Física, Universidad Técnica Federico Santa María, Avenica Vicuña Mackenna 3939, San Joaquín, Santiago, Chile\\
$^{8}$N\'ucleo de Astrof\'isica, Universidade Cidade de S\~ao Paulo, Galv\~ao Bueno 868, Liberdade, 01506-000, S\~ao Paulo,SP, Brazil\\
$^{9}$School of Physics and Astronomy, University of Birmingham, Birmingham, B15 2TT, United Kingdom\\
$^{10}$INAF-Osservatorio Astronomico di Brera, via Brera 28, I-20121 Milano, Italy\\
$^{11}$Ruhr University Bochum, Faculty of Physics and Astronomy, Astronomical Institute, Universit\"atsstr.150, 44801 Bochum, Germany\\
$^{12}$Centre for Space Research, North-West University, Potchefstroom 2520, South Africa
}

\date{Accepted XXX. Received YYY; in original form ZZZ}

\pubyear{2023}

\begin{document}
\label{firstpage}
\pagerange{\pageref{firstpage}--\pageref{lastpage}}
\maketitle

\begin{abstract}
Theoretical and observational studies have suggested that ram-pressure stripping by the intracluster medium can be enhanced during cluster interactions, boosting the formation of the "jellyfish" galaxies. In this work, we study the incidence of galaxies undergoing ram-pressure stripping in 52 clusters of different dynamical states. We use optical data from the WINGS/OmegaWINGS surveys and archival X-ray data to characterise the dynamical state of our cluster sample, applying eight different proxies. We then compute the number of ram-pressure stripping candidates relative to the infalling population of blue late-type galaxies within a fixed circular aperture in each cluster. We find no clear correlation between the fractions of ram-pressure stripping candidates and the different cluster dynamical state proxies considered. These fractions also show no apparent correlation with cluster mass. To construct a dynamical state classification closer to a merging "sequence", we perform a visual classification of the dynamical states of the clusters, combining information available in optical, X-ray, and radio wavelengths. We find a mild increase in the RPS fraction in interacting clusters with respect to all other classes (including post-mergers). This mild enhancement could hint at a short-lived enhanced ram-pressure stripping in ongoing cluster mergers. However, our results are not statistically significant due to the low galaxy numbers. We note this is the first homogeneous attempt to quantify the effect of cluster dynamical state on ram-pressure stripping using a large cluster sample, but even larger (especially wider) multi-wavelength surveys are needed to confirm the results.

\end{abstract}


%%%%%%%%%%%%%%%%%%%%%%%%%%%%%%%%%%%%%%%%%%%%%%%%%%

%%%%%%%%%%%%%%%%% BODY OF PAPER %%%%%%%%%%%%%%%%%%

\section{Introduction}
\label{sec:Introduction} 

Galaxies are distributed in different environments through clusters, groups, filaments, and voids glowing in the cosmic web with the most diverse shapes. Elliptical galaxies are more likely to be found in high-density environments, while spirals are more commonly found in low-density regions \citep{Dressler1980}. This observed dichotomy triggered the search for the physical mechanisms responsible for causing it. 

Nowadays, we know a large number of gravitational and hydrodynamical phenomena that can influence galaxy evolution \citep[see][for a review]{Boselli2006}. The primary gravitational processes are galaxy harassment \citep[accumulation of successive fast encounters within clusters,][]{Moore1996}, galaxy-galaxy interactions \citep{Spitzer1951}, and galaxy-cluster interactions \citep{Byrd1990}. These mechanisms act on all galactic components, particularly gas, stars, and dark matter. On the other hand, hydrodynamical processes such as thermal evaporation \citep{Cowie1977}, viscous stripping \citep{Nulsen1982}, and ram-pressure stripping \citep[RPS,][]{Gunn1972} affect directly only the gas component. In addition, starvation \citep{Larson1980} and pre-processing \citep{Fujita2004} combine gravitational and hydrodynamical mechanisms. Among all of those mentioned physical processes, RPS has been proven to be one of the most efficient for impacting galaxy evolution inside clusters \citep{Boselli2021}.

In particular, it has been shown that significant gas removal occurs on the first infall and that RPS is more efficient in massive clusters where it can start stripping galaxies at larger cluster-centric radii %than in low-mass systems
\citep{Jaffe_2015,Jaffe2018,Oman_2021}. Also, RPS can enhance star formation (SF) for the galaxies' stellar mass, as seen on the SF main sequence \citep{Bekki2009,Vulcani2018} and produce optically bright tails of stripped material, which give rise to jellyfish-like features \citep{Smith2010, Ebeling2014,Fumagalli2014,Poggianti_2017}.

Models of RPS often assume simplified virialised and symmetric clusters. However, in the hierarchical scenario, clusters of galaxies continue to grow from the accretion of small groups and even from collisions with structures of comparable size \citep{PressSchechter1974}. Cosmological simulations indeed show that massive clusters ($\sim$10$^{15} h^{-1} {\rm M}_{\odot}$) in the local Universe have accreted $\sim$40\,per cent of their galaxies through groups more massive than $10^{13} h^{-1} {\rm M}_{\odot}$ \citep{McGee_2009}. In addition, observations suggest that $10$--$20$\,per cent of clusters at $z < 0.3$ are undergoing mergers with other clusters \citep{Katayama_2003,Sanderson_2009,Hudson_2010}. Major cluster-cluster mergers are one of the most energetic phenomena since the big bang and can release $\gtrsim$ $10^{64}$\,ergs of energy during one crossing time ($\sim$ 1\,Gyr) \citep[][]{Sarazin2002}. 

The idea that highly energetic phenomena, such as the accretion of large galaxy groups or collisions between galaxy clusters, exert a significant impact on the galaxy evolution has been referred to as "post-processing" \citep{Vijayaraghavan2013}. This idea has been supported by observational studies that found quenching of the SF \citep{Fujita_1999,Domainko_2006, Kapferer_2009}, enhanced SF in galaxies near shock regions of the intracluster medium (ICM) \citep{Stroe2014,Stroe2015}, both quenching and enhancement of the SF \citep{Hwang2009,Ma2010,Stroe2014,Kelkar_2020}, and enhancement of RPS in dynamically disturbed clusters \citep{Owers2012,Rawle2014, McPartland2016,Ebeling2019,RomanOliveira2019,Ruggiero2019}. Similarly, hydro-dynamical simulations show that cluster mergers can increase the RPS through higher ICM densities or relative velocities between the galaxies and the ICM \citep{Vijayaraghavan2013,McPartland2016,Ruggiero2019}. 
These apparently contradictory results show that the impact of cluster growth on galaxy evolution is not fully understood. 

One of the main difficulties in galaxy evolution studies in disturbed clusters is defining, robustly, the dynamical state of large samples of clusters in a way that the time sequence and magnitude of the disturbances are portrayed. There are many substructure diagnostics for galaxy clusters at different wavelengths, each reflecting a different aspect of a cluster's dynamical state. For instance, the X-ray emission from the ICM has been widely used to study merger-induced shocks \citep{Markevitch2007,Botteon2018,Ha2018} and cold fronts \citep[CF,][]{Owers_2009,Ascasibar2006}. These are isobaric surface brightness discontinuities where the denser regions are colder than the rarefied ones and provide evidence of gas motion even in clusters that suffered minor mergers \citep{Ghizzardi2010,Birnboim2010,Hallman2010}.

On the other hand, if we look at clusters in the optical, we can often find substructures traced by the cluster galaxies themselves, which can share common positions and velocities. Optical substructures are indeed recently accreted halos that can lose $25$--$45$\ per cent of their mass per pericentric passage \citep{Taylor2004}, getting mostly disrupted within 1--3\,Gyr after infall \citep{Choque2019,Benavides2020}.

Because of the collisional nature of the ICM, the X-ray performs better than the optical in tracing interactions for longer. \citet{Poole2006} showed, using a simulated {\it Chandra} image of a cluster system, that X-ray morphological parameters and offset between the X-ray peak and centre of mass can be identified up to the second pericentric passage at $z = 0.1$, and significant temperature fluctuations can persist even after the system reaches virialisation. Moderate and massive mergers last for 4.5--5.5\,Gyr, while minor mergers last for up to $\sim$ 2\,Gyr longer in their set of simulations. The intrinsic characteristics of each dynamical state method, such as the timescale of relaxation, require that more than one diagnostics be used in order to obtain a robust classification of a large sample. 

In this paper, we perform the largest and most homogeneous observational study to date of the impact of cluster dynamical state on RPS. 
We do this by computing the incidence of RPS candidates (relative to the infalling population of cluster galaxies) as a function of a variety of proxies for cluster dynamical state for a large sample of nearby clusters hosting RPS candidates from \citet[P16 hereafter]{Poggianti2016}. \citetalias{Poggianti2016} is the largest and most homogeneous sample of optically selected RPS candidates in the nearby Universe. The importance of RPS in local clusters was recently investigated in \citet[V22 hereafter]{Vulcani2022}, by characterising the fraction of galaxies with optical ram-pressure features using the same data-set. 
Our study expands \citetalias{Vulcani2022} by comparing the RPS candidate fractions against cluster dynamical state and also considering several important caveats.

This paper has the following structure. In Sec.~\ref{sec:sample}, we present our cluster and galaxy sample. In Sec.~\ref{sec:cl_dyn_stage}, we present the different methods used to robustly classify the clusters' dynamical states. In Sec.~\ref{sec:jelly_incidence}, we compare the fractions of RPS candidates in different apertures against cluster dynamical state, and finally, the discussions and conclusions of our analysis are given in Sec.~\ref{sec:discussion} and \ref{sec:conclusions}. 

Throughout this paper, we adopt a $\Lambda$CDM cosmology with $H_{0} = 70\,{\rm km}\,{\rm s}^{-1}\,{\rm Mpc}^{-1}$, $\Omega_{\rm M} = 0.3$, $\Omega_{\Lambda} = 0.7$.

%%%%%%%%%%%%%%%%%%%%%%%%%%%%%%%%%%%%%%%%%%%%
%%%%%%%%%%%%%%%%%%%%%%%%%%%%%%%%%%%%%%%%%%%%
\section{Data and samples}
\label{sec:sample}
Since this study attempts to statistically address the incidence of RPS candidates in clusters of different dynamical states, we based our analysis on the largest and most homogeneous sample of optically-selected RPS candidates at low redshift known to date \citepalias[][described in Sec.~\ref{sec:jellyfish}]{Poggianti2016}. This sample of stripped galaxies is ideal, as it is hosted by a wide variety of clusters with homogeneous optical data from the WIde-field Nearby Galaxy Cluster Survey (WINGS) and OmegaWINGS (Sec.~\ref{sec:optical}), some of which are also covered by archival {\it Chandra} and \textit{XMM-Newton} X-ray data and radio observations (Sec.~\ref{sec:Xrays} and \ref{sec:Radio}), as described in the following.

\subsection{Cluster sample}
\label{sec:optical} 

WINGS \citep{Fasano_2006} is a multi-wavelength imaging survey primarily in $B$ and $V$ bands \citep{Varela_2009} of 77 X-ray selected clusters of galaxies in the local Universe \citep{Ebeling_1996,Ebeling_1998,Ebeling_2000} limited only on the distance from the galactic plane ($b \geqslant 20^{\circ}$) and in redshift ($z = 0.04 - 0.07$). A multi-object spectroscopic follow-up obtained data for galaxies in 48 of those clusters \citep{Cava_2009, Moretti2014}. The WINGS sample covers a wide range of masses $M_{200}$ ($\sim$ 5 $\times$ $10^{14}$ to > $10^{15}$ $M_{\odot}$), X-ray luminosities (log$L_{\rm X}$ = $43.2$ -- $44.7$, measured in the $0.1$ -- $2.4$\,keV energy band) and has a field of view (FOV) of $34' \times 34'$. When observed with good seeing, most of the clusters reached a magnitude $V \sim$ 22.0 mag or even fainter than this. The OmegaWINGS survey extended the coverage area to $1^{\circ} \times 1^{\circ}$ of each cluster, for 46 clusters \citep{Gullieuszik_2015} from which 33 also had spectroscopic follow-up \citep{Moretti2017}.

The physical radius $R_{200}$ is taken from \citet{Biviano_2017} when available; if not, they were obtained by following the method used in \citeauthor{Durret2021}, (\citeyear{Durret2021}; see also \citealt{Kolcu2022}), inserting the observed line of sight velocity dispersion in ${\sigma_{\rm cl}} = 1090 \times \left [{h(z) \times M_{200}} \right ]^{1/3}$, Eq. 1 from \citet{Munari_2013} and the relation

\begin{equation}
R_{200} = \bigg(\frac {G\; M_{200}} {100\; H^{2}_{z}}\bigg)^{1/3},
\end{equation}

where the Hubble constant at $z$ is $H^{2}(z) = H^{2}_{0}[\Omega_{0}(1 + z)^{3} + \Omega_{R}(1 + z)^{2} + \Omega_{\Lambda}]$ and $h(z) = H(z)/100$ \citep{Peebles1993}. $M_{200}$ is expressed in units of $10^{15}$ $M_{\odot}$, and $\sigma_{\rm cl}$ is the uni-dimensional velocity dispersion in units of km/s.

In total, WINGS and OmegaWINGS have obtained 24122 galaxy redshifts. After a recent systematic search for additional redshifts in the photometric sample, \citetalias{Vulcani2022} and references therein increased the spectroscopic sample to 46700. The additional redshifts were considered for assigning cluster membership and for the substructure analysis (see Sec.~\ref{sec:ds}). The cluster members were identified based on a $3\sigma$ clipping method as described in \citet{Paccagnella_2017}.

\subsubsection{P16 cluster sample}
\label{sec:clusters} 

The parent cluster sample comprises 72 WINGS clusters from \citetalias{Poggianti2016} (41 are also in OmegaWINGS), which presented the first sizeable systematic search for RPS candidates at low redshift. Details about the galaxies in this sample can be found in Sec.~\ref{sec:galaxies}

The spatial coverage of the clusters in the WINGS/OmegaWINGS surveys varies as a result of the different cluster sizes and redshifts and the different FOV of the surveys. To obtain a physical radius that maximised the size of our cluster sample and its radial extent, we calculated the maximum radii for which each cluster is fully covered by the FOV of the WINGS and/or OmegaWINGS surveys, finding a range from 0.3 $R_{200}$ to 2.1 $R_{200}$. Ideally, we would like to study galaxies well beyond $R_{200}$ to understand how the different dynamical states impact the incidence of RPS candidates in the infalling region. However, this is not always possible, and limiting the analysis to clusters that satisfy this criterion would limit the sample significantly. We found that 75\,per cent of the sample have spatial coverage reaching $0.7\,R_{200}$ or beyond. This radial cut limits our sample to 52 clusters from \citetalias{Poggianti2016} in the different dynamical states without penalising the sample size significantly. We, therefore, use these 52 clusters in our analysis.

\subsubsection{X-ray sample}
\label{sec:Xrays} 

We cross-matched the \citetalias{Poggianti2016} sample with a catalogue of 964 galaxy clusters from {\it Chandra} with X-ray morphological parameters computed by \citet[Y20 hereafter]{Yuan2020}. Subsequently, we replicated the aforementioned process with the sample of a complementary study for 1308 clusters from \textit{XMM-Newton} performed by \citet[Y22 hereafter]{Yuan2022}. Analysing the correlation between these parameters helps to build the widely used dynamical state diagnostics presented in Sec.~\ref{sec:cl_dyn_stage}. We found 35 {\it Chandra} and 36 \textit{XMM-Newton} clusters that have complete photometric coverage at $0.7\,R_{200}$ and are in common with \citetalias{Poggianti2016}, for a total of 45 distinct objects. 

As a sanity check, we compared the X-ray morphological parameters obtained in \citetalias{Yuan2020} with the ones obtained in \citetalias{Yuan2022} for the 26 objects in common between the two data-sets. Overall we found a small departure from the reference 1:1 line and a high Spearman correlation coefficient ($\geq 0.7$), which agrees with \citetalias{Yuan2022}, who run a similar comparison with a larger cluster sample of 351 clusters. Therefore, we decided to merge both catalogues. We preferentially used the {\it Chandra} data since their spatial resolution is better than the \textit{XMM-Newton}. This feature is very important to the dynamical state classification of the cluster sample.

\subsubsection{Spectroscopically complete sample}
\label{sec:Spec_cmpl} 

Out of 52 \citetalias{Poggianti2016} clusters that reached $0.7\,R_{200}$, 49 have spectroscopic completeness $> 50$\,per cent at a total magnitude brighter than $V = 17.77 \, \rm mag$. We refer to these clusters as the spectroscopically complete sample. This magnitude limit comes from the photometric analysis performed in Sec.~\ref{sec:phot_frac}, where we correct the RPS candidate fractions for the field contamination. It was necessary to use this limit so that we could compare both samples, photometric and spectroscopic, fairly.

\begin{figure}
	\includegraphics[width=\columnwidth]{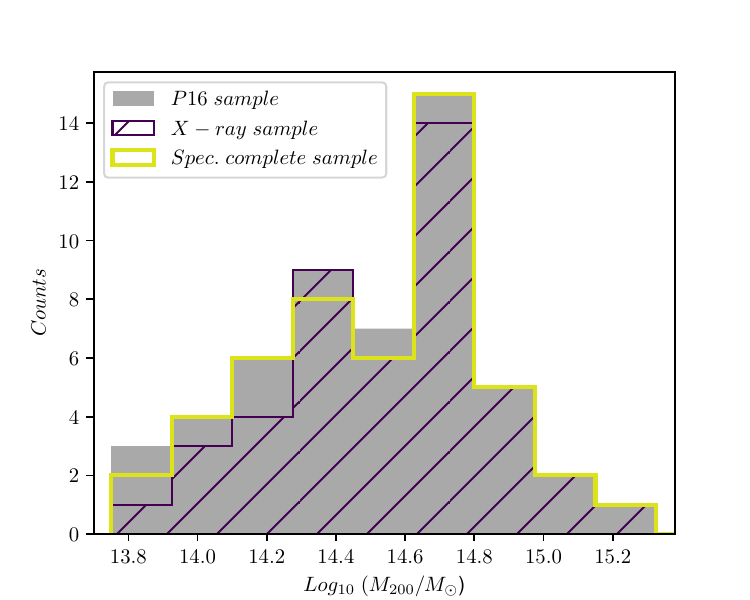}
    \caption{Mass distribution of the different cluster samples used in this paper. Masses were retrieved from \citet{Biviano_2017}. The parent sample is from \citetalias{Poggianti2016} (52 clusters), followed by the cluster samples with available X-ray data (45 clusters), and the spectroscopically complete cluster sample (49 clusters). Only clusters that have complete photometric coverage at $0.7\,R_{200}$ are included in those samples.}
    \label{fig:Mass_hist}
\end{figure}

Figure \ref{fig:Mass_hist} shows the halo mass distribution of the different samples used in this paper, including the \citetalias{Poggianti2016}, X-ray, and spectroscopically complete samples mentioned in this section. These masses were computed in \citet{Biviano_2017}. 
In Tab.~\ref{tab:cls_properties}, we specify to which sample(s) each cluster belongs.

\subsection{Galaxy sample}
\label{sec:galaxies} 

WINGS and OmegaWINGS provide valuable information for their galaxy samples, such as galaxy colour and morphology, as well as spectroscopic quantities, permitting the study of galaxy evolution in the local Universe in an unprecedented way with a large sample.

The total absolute magnitudes, $M_{\rm B}$ and $M_{\rm V}$, were derived from the total (SE{\footnotesize XTRACTOR} AUTO) observed $B$ and $V$ magnitudes \citep{Moretti2014, Gullieuszik_2015}, corrected for distance modulus, foreground Galaxy extinction, and k-corrected using tabulated morphology dependant values from \cite{Poggianti1997}. Morphologies were obtained by MORPHOT \citep{Fasano_2012}, an automated, non-parametric tool for the morphological-type estimate of large galaxy samples. 

In our photometric and spectroscopic analysis, we only use galaxies with $M_{V} < -19.5$, which is the magnitude limit used in MORPHOT classifications. Similar to \citet{Fasano_2012}, we consider late-type galaxies, the ones with MORPHOT classifications within $0.5 \leq Morphological\;Type \leq 10.5$. This revised Hubble type range encompasses S0/a, Sa, Sab, Sb, Sbc, Sc, Scd, Sd, Sdm, Sm, Im, and compact irregular galaxies (cI).

The BCGs used in this work are primarily those identified in \citet{Fasano_2010}. Because in this paper we use BCGs to define the cluster centre (when possible, see Sec.~\ref{sec:flowchart}), we revised case by case using the optical and X-ray data in hand and found that in a few cases, their BCGs were not the bright galaxy closest to the X-ray emission peak. We describe these cases in the appendix~\ref{sec:centres}.

\subsubsection{Ram-pressure stripping candidates}
\label{sec:jellyfish} 

In \citetalias{Poggianti2016}, the identification and classification of the RPS candidates were made by a few human classifiers with no information about the galaxy environment, cluster centre or memberships. The \citetalias{Poggianti2016} cluster sample only included clusters with $B$ and/or $V$-band seeing $\leq 1.2\, \rm arcsec$ to visually search for the RPS candidates. 
Galaxies with tails, gas disturbance, asymmetrical star formation, and bow shock features were selected as RPS candidates. The classifiers assigned a classification (Jclass) from 1 to 5 according to how confident they were that the galaxy was undergoing RPS. 
We excluded all the \citetalias{Poggianti2016} candidates that were flagged with possible gravitational interaction (with comments such as "tidal", "merger", or "harassed"), getting a final sample of 53 blue late-type RPS candidates in 52 \citetalias{Poggianti2016} clusters. We stress that clusters with no candidates within $0.7\,R_{200}$ are also included in our analysis.

As discussed in \citetalias{Poggianti2016}, most RPS candidates have low Jclass. In fact, in our sample, only $\sim 28$\,per cent of the RPS candidates have a Jclass value of 3 or higher. While it is true that lower Jclass values indicate more uncertain cases of RPS, based on GASP\footnote{GAs Stripping Phenomena in galaxies with MUSE (GASP) is an integral-field spectroscopic survey with MUSE at the VLT aimed at studying gas removal processes in galaxies \citep{Poggianti_2017}.}data of a subset of the \citetalias{Poggianti2016} RPS candidates, Poggianti et al. (in prep) show that the "success rate" of identifying ram-pressure stripped galaxies (across all Jclass values) through broad-band optical imaging in local clusters is approximate $\sim 86$\,per cent. In other words, the vast majority of the RPS candidates used in our study are expected to be "real" cases of ram-pressure stripping at play.

%%%%%%%%%%%%%%%%%%%%%%%%%%%%%%%%%%%%%%%%%%%%%%%%%%%%%%
\section{Determining clusters dynamical state}
\label{sec:cl_dyn_stage} 

 As discussed before, due to observational limitations and biases, it is very difficult to reliably quantify cluster substructures using a single proxy. For this reason, we employed all the information that was available to quantify the amount of substructure in our sample. In particular, we combined several optical and X-ray disturbance proxies available in the literature for the \citetalias{Poggianti2016} sample and additionally derived other parameters described below in order to understand how those different techniques correlate with each other. 

\subsection{Optical proxies}
\label{sec:optical_proxies}

\subsubsection{Dressler-Shectman analysis}
\label{sec:ds}
The Dressler-Schectman (DS) test \citep{DresslerShectman1988} estimates the presence of substructures within a cluster by identifying the nearest neighbours $N_{\rm nn}$ of each galaxy (confirmed spectroscopic member) and comparing the difference of their kinematics with the global kinematics of that cluster. More precisely, the test computes the differences between the velocity of each galaxy $\overline{v}_{\rm local}$, and velocity dispersion $\overline{\sigma}_{\rm local}$ with respect to the cluster mean, $\overline{v}_{\rm cl}$ and $\overline{\sigma}_{\rm cl}$. The following equation gives the deviations of the individual galaxies to the global kinematic parameters of the cluster: 

\begin{equation}
\delta_{i}^{2} =  \bigg(\frac {N_{\rm nn}+1} {\sigma_{\rm cl}^{2}}\bigg) \Big[(\overline{v}_{\rm local}^{i}-\overline{v}_{\rm cl})^{2}+(\sigma_{\rm local}^{i}-\sigma_{\rm cl})^{2}\Big]
\end{equation}

This test can be evaluated in two different ways. i) If the sum of deltas divided by the number of members exceeds 1, the cluster is said to have substructures. ii) The result obtained in i) can be further investigated by using Monte Carlo re-sampling of the local velocity values to obtain the probability that the cluster has a substructure \citep[e.g.][]{Jaffe_2013}. However, these methods do not identify which galaxies are members of the substructures.

\citet{Biviano_2017} introduced an improved version of the \texttt{DS} test that is capable of identifying which galaxies belong to a substructure, the \texttt{DS+} technique. This method is fully described in \citet{Biviano_2017} and tested with numerical simulations in \citet{Benavides2023}. Here we briefly summarise it. \texttt{DS+} does not restrict $N{\rm nn}$ to a fixed number. Instead, it considers any possible multiplicity of neighbouring galaxies and checks for differences in their kinematics from that of the cluster as a whole. After assigning a significance to each detected group using Monte Carlo re-sampling, the groups that share one or more galaxies with a more significant group are disregarded to avoid overlapping. Groups that are close enough in distance and velocity are merged to avoid fragmentation \citep[see eq.\ (3) of][]{Benavides2023}. In addition, the \texttt{DS+} method differs from the \texttt{DS} method in that it considers the deviations in the first and second moments of the velocity distributions separately. Since real groups are not expected to have velocity dispersion larger than clusters, \texttt{DS+} does not consider as significant positive deviations of the velocity dispersion of the type $\sigma_{\rm local}^i > \sigma_{\rm cl}$. Moreover, \texttt{DS+} encompasses $\sigma_{\rm cl}(R)$, i.e. the cluster velocity dispersion profile (VDP), and not the whole cluster velocity dispersion, as reference. Two choices are possible for the VDP: one is obtained by smoothing the observed cluster VDP with the LOWESS (LWS) technique \citep{Gebhardt1994}, the other is based on the Navarro Frenk White (NFW) mass profile \citep{Navarro1997}, as described in \citet{Biviano_2017}.

We evaluate two global substructure indicators based on preliminary work where the authors applied \texttt{DS+} on simulated data \citep{Benavides2023}. One is $f_{\rm sub}$, the fraction of members that belong to sub-clusters, $N_{\rm sub}$, with respect to the number of members in the whole cluster, $N_{\rm mem}$. The other is $f_{\rm max}$, the fraction of members that belong to the richest sub-cluster, $N_{\rm max}$, with respect to $N_{\rm mem}$. We have values for each of \begin{math}f_{\rm sub}= \frac {N_{\rm sub}} {N_{\rm mem}}\end{math} and \begin{math}f_{\rm max} = \frac {N_{\rm max}} {N_{\rm mem}}\end{math}.

We chose to present in this work only the results obtained with the LWS VDP since using this profile $f_{\rm sub}$ and $f_{\rm max}$ correlate slightly better than when using the \citet{Navarro1997} mass profile. 

\subsubsection{Magnitude Gap}
During the growth of a halo, the central galaxy undergoes several mergers, cannibalising the satellite galaxies \citep[e.g.][]{Contreras-Santos2022}. Over time, the difference between the mass of the central and satellite galaxies increases, which also expresses itself in a greater difference in magnitude. The time a halo was formed is linked to its dynamical state. Dynamically evolved clusters tend to have a large magnitude gap between their first and second-ranked galaxies. Hence, the magnitude gap has been widely used in the literature to segregate between dynamical states \citep[e.g.][]{Dariush2007,Gozaliasl2014,Raouf2019}. In particular, a magnitude gap $\geq 2$ has traditionally been used to distinguish a special class of very passive systems where no significant interaction happened in their recent history. These systems are known as fossil groups \citep{Jones_2003}.

Following the method described by \citet{Dariush2007}, we searched for the two brightest galaxies in the absolute V magnitude within a radius of $0.5\,R_{200}$ around the brightest galaxies of each cluster identified in \citet{Fasano_2010}. Late-type galaxies and non-members (based on available spectroscopic redshifts, when available) were excluded from this search. The magnitude gap $\Delta M_{1,2}$ is obtained with the following equation,
\begin{equation}
\Delta M_{1,2}= {{V}_{\rm rank2}-{V}_{\rm rank1}}
\end{equation}

where ${V}_{\rm rank1}$ and ${V}_{\rm rank2}$ are the absolute V magnitudes of the first and second-ranked galaxies, respectively.

\subsection{X-ray proxies}
\label{sec:xray_proxies} 

Here we present several X-ray morphological parameters computed in \citetalias{Yuan2020} and in \citetalias{Yuan2022} that have been extensively used in the literature to classify the dynamical state of galaxy clusters. Inspired by traditional optical parameters \citep[see][]{Conselice2003}, these X-ray proxies include concentration (how centrally concentrated the X-ray emission is), centroid shift (how much the centroid moves as the aperture size changes), asymmetry (degree of asymmetry in the cluster's surface brightness distribution), and power ratio (how non-uniform mass distribution within the cluster is).

These X-ray morphological parameters offer valuable insights into the dynamical state of galaxy clusters. Relaxed clusters are usually more concentrated, and less asymmetric, with smaller power ratios and centroid shifts. For more detailed information on the morphological parameters computation, refer to \citetalias{Yuan2020} and \citetalias{Yuan2022}.

\subsubsection{Cold fronts}
{\it Chandra} sub-arcsecond angular resolution enabled the first observations of cold fronts \citep{Markevitch_2000}. CFs are sharp discontinuities observed in the X-ray surface brightness maps \citep[see][for a review]{Markevitch2007}, where the denser side of the front is colder. The pressure remains nearly constant in a CF. Across the CF, a jump in gas metallicity has been seen in multiple instances, possibly due to enriched low entropy gas being stripped by ram-pressure from satellite galaxies in the rich environment \citep{Markevitch_2000,Simionescu2010,Ghizzardi2013}. CFs are mainly grouped into two classes according to their formation mechanism, merger-remnant cold fronts (MCF) and sloshing cold fronts (SCF). The collision of two shocks or gas streams can provide an alternative way of forming a cold front \citep{Zinger2018}. MCFs form at the leading edge of the substructure due to ram-pressure confinement of the infalling substructure when a cold and dense substructure dives through a hotter and more diffuse medium \citep{Markevitch_2000,Vikhlinin2001}. SCFs occur when an off-axis minor merger disturbs the cool core of the main cluster. The disturber inserts angular momentum into the main cluster while crossing it, causing its internal configuration to "slosh" sub-sonically around the gravitational potential. This disturbance creates spiral-like patterns around the cluster core, extending to large radii \citep{Tittley2005,Ascasibar2006,Lagana2010,Roediger2011,ZuHone2011,Vaezzadeh2022}. CFs play a significant part in the overall ICM dynamics because of the variety of formation methods and their widespread nature in the ICM. We searched for evidence of CF in the literature when available, as shown in Tab.~\ref{tab:cls_properties}.

\subsection{Radio proxies}
\label{sec:Radio} 
The shock waves and turbulence induced by mergers in the ICM often manifest as diffuse radio emissions such as radio relics and halos that can be used as a further diagnostic of the dynamical state of the cluster \citep[for a review see][]{Brunetti2014,vanWeeren2019}. Radio relics are megaparsec-sized radio emissions. The shocks travelling through the ICM re-accelerate the cosmic rays (CR) in the ICM. The synchrotron emission observed in radio relics is produced by re-accelerated CRe (electrons). The turbulence injected by the merger often produces large-scale, unpolarised radio sources, known as radio halos, in the cluster centre. 

Some clusters in our sample have information in the literature about the presence of radio relics and halos. We caution that not all the clusters were observed at radio wavelengths and that the observations were carried out at different frequencies and sensitivities. Although recently, relics were observed in pre-merger clusters \citep{Gu2019,Sarkar2022}, typically, radio halos and relics are evidence of post-merger events. We list the references for the clusters with radio relics and halos in Tab.~\ref{tab:cls_properties}.
    
%%%%%%%%%%%%%%%%%%%%%%%%%%%%%%%%%%%%%%%%%%%%%%%%%%%%%%

\subsection{Combined wavelength proxies}

%\subsubsection{X-ray peak or centroid to BCG offset}
Several studies have shown that the gas disturbance can be inferred by the projected offset between the X-ray peak and the BCG, $\Delta X_{\rm BCG}$ \citep{Sanderson_2009,Katayama_2003,Hudson_2010,Lopes_2018,Zenteno2020}, since major mergers can cause offsets of dozens of kilo-parsecs \citep[$\sim$ 939\,$h_{71}^{-1}$\,kpc for A3376,][]{Hudson_2010}. Similarly, other authors have used the separation between the X-ray centroid and the BCG to infer the dynamical state of the clusters. The distance to the X-ray peak was found to be more effective for segregating major mergers, since only major mergers can cause the X-ray peak to dissociate considerably from the BCG \citep{DeLuca2021}. On the other hand, according to \citet{Mann_ebeling2012}, using the X-ray centroid/peak does not change the results significantly.
\citet{Rasia2013} and \citet{DeLuca2021} found that a combination of X-ray morphology with dynamical parameters gives a better inference of the cluster dynamical state. $\Delta X_{\rm BCG}$ suffers from projection effects and is most sensitive to plane-of-the-sky mergers, while dynamical proxies are best for line-of-sight (LOS) interactions.

\subsection{Correlation between dynamical state proxies}
\begin{figure*}
    \includegraphics[width=17.8cm]{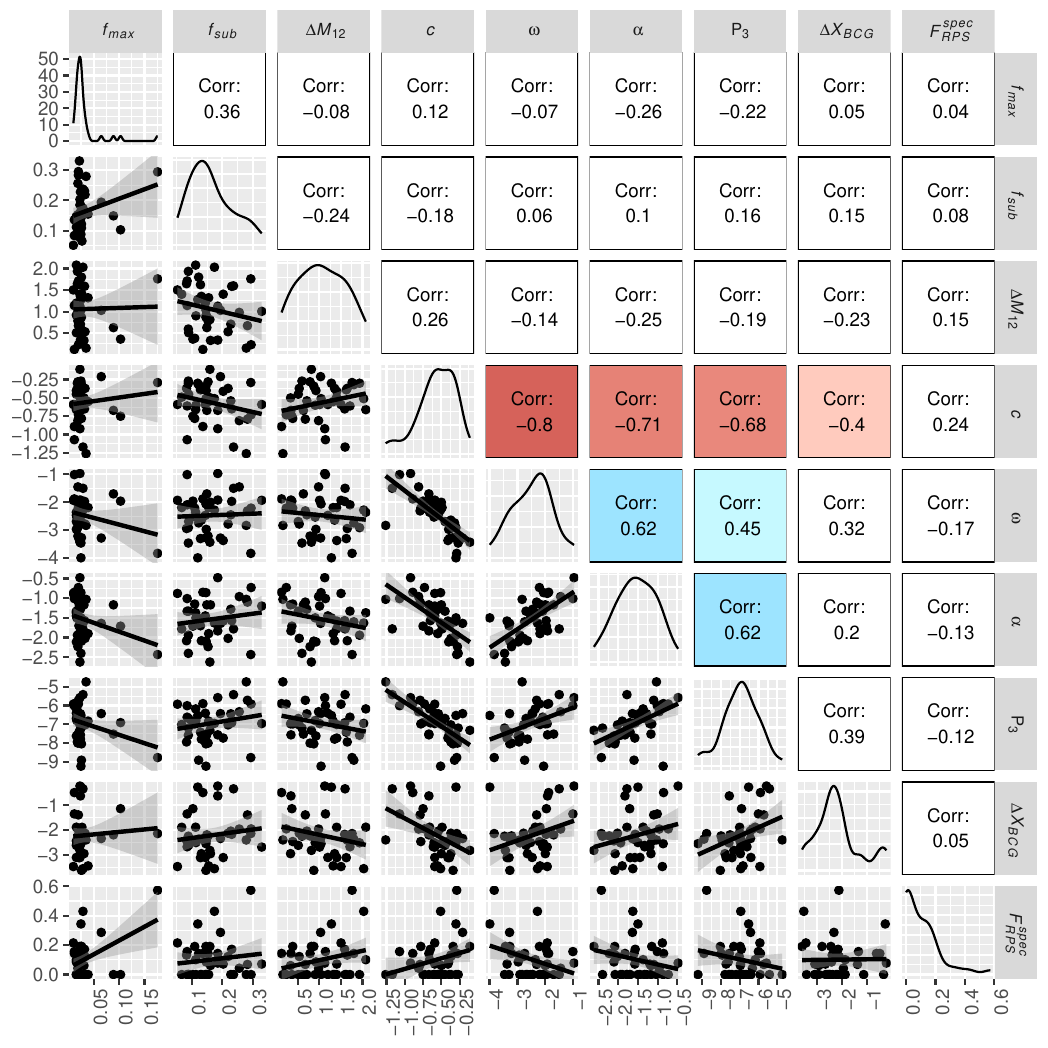}
    \caption{Correlation matrix of the main dynamical state diagnostics used in this work and the fractions of ram-pressure candidates within a $0.7\,R_{200}$ radius. The different proxies are explained in Sec.~\ref{sec:cl_dyn_stage}, and the last column is discussed in Sec.~\ref{sec:phot_frac}. The panels on the top of the diagonal show the Spearman correlation coefficient. The coloured panels show the most significant correlations (blue) and anti-correlation (red). The Spearman correlation varies from -1 to 1. A strong correlation is close to 1, a strong anti-correlation is close to -1, and values around 0 do not correlate. The panels on the diagonal show the distributions of each proxy, while the ones on the bottom of the diagonal show the bi-variate scatter plots (grey dots show each cluster in our sample) with linear fits (black lines). There is no clear separation between relaxed and disturbed clusters. RPS fractions do not correlate with the dynamical state proxies used in this work.}
    \label{fig:cl_dyn_stage}
\end{figure*}

In Fig.~\ref{fig:cl_dyn_stage}, we present the Spearman correlation matrix of the main (optical and X-ray) dynamical state diagnostics used in this work and the fractions of infalling galaxies (blue late-type) undergoing RPS within a $0.7\,R_{200}$ aperture (which will be discussed in Sec.~\ref{sec:jelly_incidence}). The correlations between each pair of diagnostics are given on the top of the matrix diagonal. On the bottom of the diagonal, we see the bi-variate scatter plots with the linear model fitted lines. The distribution of each variable is shown on the diagonal as density plots. After running a Gaussian mixture model for the dynamical state parameters, we found that except for $f_{\rm max}$ and the offset between $\Delta X_{\rm BCG}$, which are bi-modal, the dynamical state proxies of our sample of clusters are uni-modal. This agrees with the results from \citet{Campitiello2022} for the distribution of the X-ray morphological parameters of 118 clusters observed with \textit{XMM-Newton}. Overall, most of the clusters present intermediate values for the different diagnostics. Only a few clusters are very relaxed or very disturbed. Perfect correlations or anti-correlations are those close to 1 or -1, respectively. The values close to zero do not correlate. We coloured the strongest correlation (blue) and anti-correlation coefficients (red).

According to the results shown in Fig.~\ref{fig:cl_dyn_stage}, the parameters that better correlate are the concentration (c) and the centroid shift ($\omega$) (-0.80), followed by the concentration and the asymmetry ($\alpha$) (-0.71), and the concentration and power ratio ($P_{3}$) (-0.68), all of which are derived from X-ray. 
 
The optical parameters do not have strong correlations between them.$f_{\rm sub}$ weakly anti-correlates with $\Delta M_{1,2}$ (-0.24) and exhibits a marginally stronger correlation with $f_{\rm max}$ (0.36). This apparent correlation is driven by an outlier, though. \citet{Roberts_2018} compared X-ray morphology parameters (asymmetry and centroid shift) with optical dynamical state diagnostics (Anderson-Darling statistic, stellar mass gap between the first and second most massive galaxies, and offset between the most massive cluster galaxy and the luminosity-weighted centre). They found a strong correlation between the centroid shift and the stellar mass gap in their cluster sample. We can compare our results to theirs by assuming that the magnitude gap can be used as a proxy for the mass gap. Their results are at odds with ours since we found a very weak anti-correlation between centroid shift and magnitude gap (-0.14). This can be due to the fact that their sample covers a much wider redshift range than ours.

There are many obstacles in identifying dynamical states using the most diverse techniques presented here. The in-homogeneity in clusters that have good quality data in the same X-ray telescope, such as {\it Chandra} and \textit{XMM-Newton}, considerably reduces the sub-sample of clusters for which we can make the comparison shown in Fig.~\ref{fig:cl_dyn_stage}. Other factors that impacted the size of this sub-sample were the lack of photometric and mainly spectroscopic coverage for large fractions of a cluster's $R_{200}$ and the spectroscopic completeness. 

\subsection{Defining a dynamical state sequence}
\label{sec:flowchart}

\begin{figure*}
	\includegraphics[width=17.8cm]{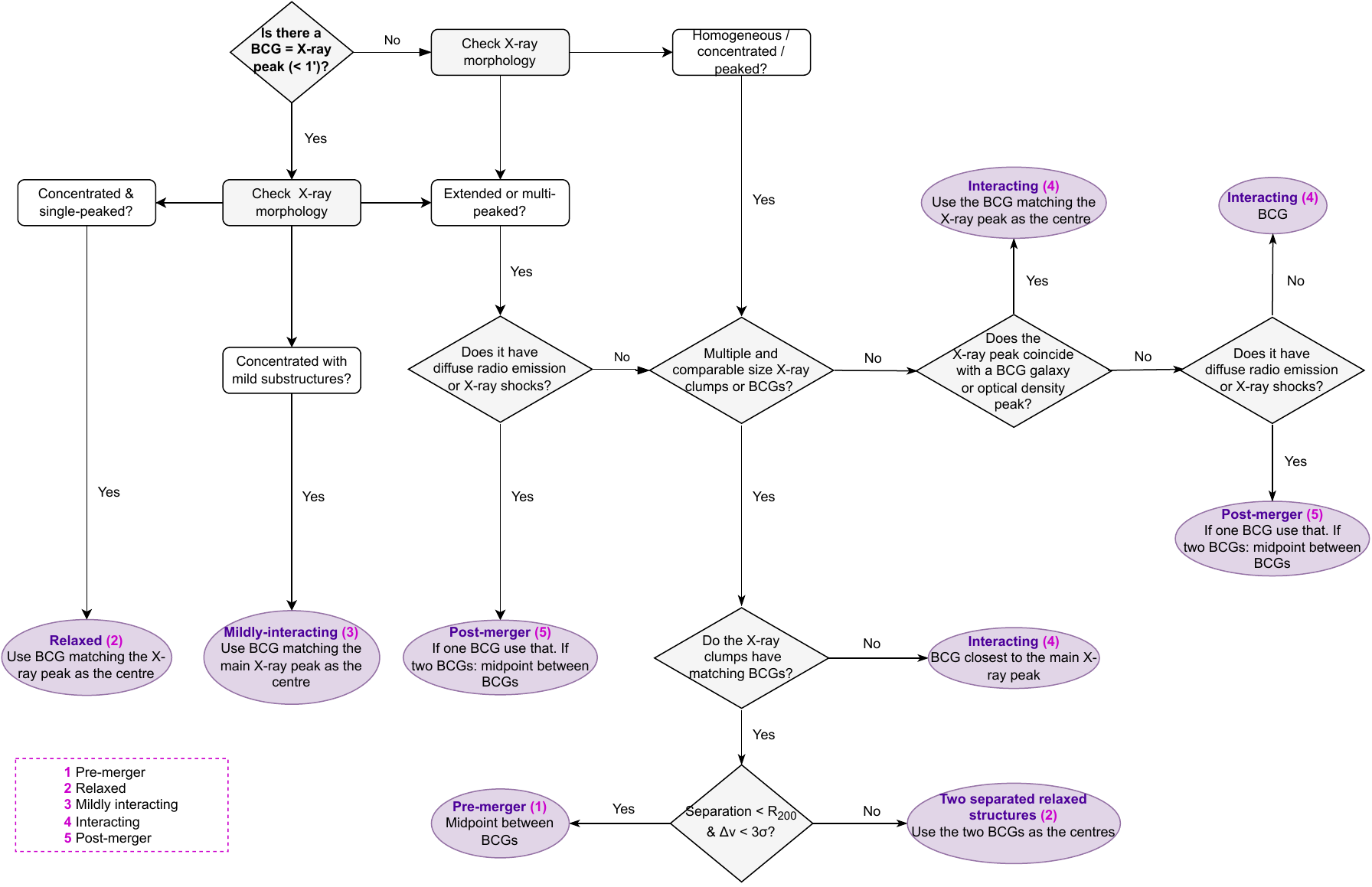}
    \caption{Flowchart used to tentatively classify the dynamical state sequence and define cluster centres. This classification was only done for clusters with X-ray data.}
    \label{fig:flow_chart}
\end{figure*}

In the previous section, we saw how different diagnostics for cluster dynamical states could yield different results, with the X-ray proxies being more consistent for this sample. In this section, we take a closer look at all the available data for the X-ray sample to define a discrete combined classification of the clusters that tentatively reflects the time sequence of the interaction and/or its intensity. At the same time, we use this classification to define, for each case, the centre to consider in our analysis (Sec.~\ref{sec:jelly_incidence}). To this end, we constructed a flowchart, shown in Fig.~\ref{fig:flow_chart}, which yields the five different possibilities for the cluster's dynamical states and centres described below. The dynamical sequence is done by comparing the positions of the BCG with the X-ray peak and checking whether there are secondary X-ray peaks and if there are radio relics seen. If multiple X-ray structures are visible, the brightest galaxy in each structure is considered a BCG, leading to multiple "BCGs" in a single cluster.

\begin{enumerate}[wide,labelwidth=!,itemindent=!,labelindent=0pt, leftmargin=0em,label={\arabic*.}]

    \item \textbf{Pre-merger}: This class of clusters has multiple and comparable size X-ray clumps, each with a matching BCG (within $1'$). We based the choice of this radius on the fact that only four BCGs in our sample have an effective radius greater than $1'$. In our pre-merger classification, the X-ray clumps are separated by less than $R_{200}$ and $3\sigma_{\rm cl}$ in velocities (using the most massive cluster radius and velocity dispersion as reference). Usually, these pre-merging clusters do not present diffuse radio emissions, although radio shocks have been observed in the pre-merger state of the cluster pair 1E 2216.0-0401/1E 2215.7-0404 \citep{Gu2019} and recently in Abell 98 \citep{Sarkar2022}. The motivation for not including these clusters in the interacting category comes from the fact that we are interested in investigating whether pre-merging environments can enhance RPS candidate fractions. We use the midpoint between the two BCGs as the centre of the pre-merger systems in our analysis as an attempt to measure the enhancement caused by the ICM compression between two clumps.
    \item \textbf{Relaxed}: The X-ray surface brightness map is concentrated, single-peaked, regular, and the X-ray peak coincides with the BCG (within $1'$). In this case, we use the BCG as the centre. There is a second case of relaxed clusters where the BCG and the X-ray peak coincide within the $1'$ criteria, but the X-ray morphology shows more than one clump of comparable sizes, with BCGs separated by more than $R_{200}$ and/or velocity differences larger than $3\sigma_{\rm cl}$. In these cases, we treat them as two separate relaxed structures and use the BCG of each one as their centres.
    \item \textbf{Mildly-interacting}: The X-ray surface brightness morphology of the cluster is still concentrated, but there is a mild amount of optical and/or X-ray substructures (sub-clumps) or disturbances (such as an X-ray spiral/sloshing pattern in the core). Since the interactions are usually small in those cases, the BCG and main X-ray peak are within $1'$ of separation, and we use the BCG as the centre.
    \item \textbf{Interacting}: When the X-ray emission of the cluster is extended, distorted or multi-peaked (but clumps do not have a clear separation), we classify the cluster as interacting and use the BCG as the centre. When a cluster has multiple comparable X-ray clumps and no matching BCGs (within $1'$ of the X-ray peak) for those clumps, we classify it as interacting and use the central bright galaxy near the main peak as the centre.
    \item \textbf{Post-merger}: If the cluster is undergoing a significant merger and also shows the presence of X-ray shocks or diffuse radio emission (relics or halo), we consider it a post-merger. In this case, we use the midpoint between the BCGs as the centre, similar to what was done for the post-merger system Abell 3376 by \citet{Kelkar_2020}. 
\end{enumerate}

\begin{figure*}
	\includegraphics[width=17.8cm]{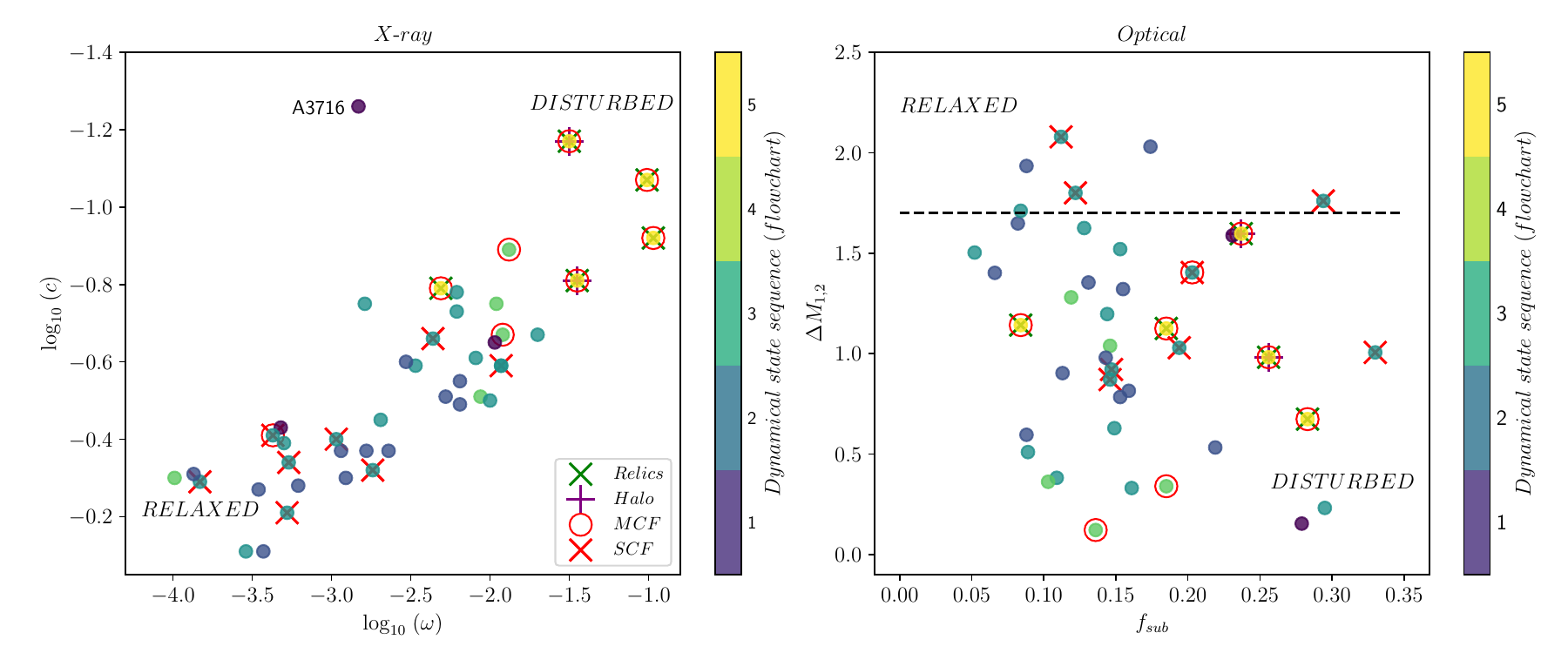}
    \caption{Left: Concentration vs. centroid shift of clusters that are in the X-ray sample coloured by our dynamical state sequence defined in the flowchart. Disturbed clusters have large centroid shifts and small concentrations. The disturbance of the clusters increases from the bottom left quadrant to the top right quadrant. Right: Magnitude gap vs. $f_{\rm sub}$ also coloured according to the dynamical state sequence from the flowchart. The black dashed line is the \citet{Raouf2019} criteria to segregate the relaxed clusters above it. In both panels, disturbed clusters have a small magnitude gap and large $f_{\rm sub}$. Green and red "X" shows radio relics and SCFs, respectively. Purple crosses show the clusters with radio halos. The red open circles show the clusters with MCFs.}
    \label{fig:Yuan_flow_chart}
\end{figure*}

In Fig.~\ref{fig:Yuan_flow_chart}, we plot the X-ray (left) and optical (right) dynamical state diagnostics with the best correlation coefficients, colour-coded by the visual dynamical state sequence performed in this section. Clusters with literature information (see Tab.~\ref{tab:cls_properties}) for radio relics, halos, sloshing, and merger cold fronts are highlighted. %Tab.~\ref{tab:cls_properties} lists the references for these features.
Note that in the right-hand panel (optical proxies), there are more clusters than in Fig.~\ref{fig:cl_dyn_stage}, where we could only use clusters that were simultaneously in the X-ray and spectroscopically complete samples to be able to directly compare against X-ray proxies\footnote{Note that adding the missing clusters in Fig.~\ref{fig:cl_dyn_stage} only causes a minor change in the distribution shape and correlation coefficients for the optical proxies.}. Overall, Fig.~\ref{fig:Yuan_flow_chart} shows that our flowchart dynamical state sequence is in best agreement with the X-ray diagnostics. This is expected since the flowchart was mainly based on some X-ray-dependent criteria, but it also adds optical information to this kind of diagnostic, making explicit, for example, that the cluster A3716, an outlier in the X-ray diagnostics plot, is a pre-merger cluster.

The most disturbed clusters should be naturally found in the upper right corner of the left-hand plot of Fig.~\ref{fig:Yuan_flow_chart}, with high X-ray centroid shift and lower concentration. In fact, \citet{Rasia2013} found that radio relics/halos tend to be located in this quadrant, which is something our sample shows as well. Another interesting feature is that clusters with SCF fall in the lower-left "relaxed" quadrant of this plot. This is not surprising when considering that SCF are caused by very minor interactions relative to a cluster merger. In the right-hand panel of Fig.~\ref{fig:Yuan_flow_chart}, the situation is less clear, as the optical proxies for cluster dynamical states are not as well correlated with each other nor with the flowchart dynamical state sequence. Clusters with radio relics prefer regions with smaller magnitude gaps (expected in cluster mergers) and those with SCF are more scattered, but with a predominant presence towards the relaxed region, as defined by the line of \citet{Raouf2019}. 

%%%%%%%%%%%%%%%%%%%%%%%%%%%%%%%%%%%%%%%%%%%%%%%%%%%%%%%

\section{Incidence of ram-pressure stripping candidates as a function of cluster dynamical state}
\label{sec:jelly_incidence} 

We are now in the position of understanding how a cluster's dynamical state influences RPS. We compute the incidence of stripped candidates in the \citetalias{Poggianti2016} sample by measuring the fraction of galaxies undergoing RPS relative to the infalling population, which we characterise as blue late-type galaxies. These gas-rich galaxies are likely the most susceptible to RPS. Also, they were previously used in \citetalias{Vulcani2022} to obtain similar RPS fractions. 

In order to define the blue galaxies, we fit the distribution of $B - V$ colour in separate $M_{V}$ bins with two Gaussians. The peak value of the redder Gaussian was taken as the midpoint of the red sequence, and the scatter was taken as the standard deviation. This was repeated across four magnitude bins, and then we used the line 1-$\sigma$ below the resultant linear fit to separate red sequence galaxies from blue galaxies. The equation of the derived line is:

\begin{equation}
\label{eq:red_seq}
(M_{B} - M_{V})= -0.037 \times M_{V}+0.046 
\end{equation}

Galaxies with a $B-V$ colour bluer than this relation are considered blue. This method is similar to that used in \citet{Crossett2017,Crossett2022}.

Furthermore, we computed the RPS candidate fractions considering two samples. We first considered the (larger) photometric galaxy sample, and to account for the fact that we do not have cluster memberships for the whole photometric sample, we performed a correction for field contamination (Sec.~\ref{sec:phot_frac}). Then, to test if the lack of cluster memberships was biasing our results, we re-did the analysis considering the (smaller) spectroscopic galaxy sample, where we can select cluster members with confidence and exclude all the non-members in the computation of the RPS candidate fraction (Sec.~\ref{sec:spec_frac}).

\subsection{Photometric RPS fractions}
\label{sec:phot_frac}

To compute the RPS candidate fractions from the photometric galaxy sample, we counted the number of RPS candidates that are blue late-type galaxies within $0.7\,R_{200}$, $N_{\rm RPS}^{\rm BSp}$ (see justification in Sec.~\ref{sec:clusters}). Then, we measured the total number of blue late-type galaxies in the same area, $N^{\rm BSp}$. To account for the field contamination, we subtracted the expected number of blue late-type galaxies in the field within the same area. The percentage of blue late-type galaxies in the field was estimated following the method used in \citet{Fasano_2015}. Similar to them, we used a comparison field sample of galaxies from the Padova-Millennium Galaxy and Group catalogue \citep[PM2GC,][]{Calvi2011,Calvi2012}, which encompasses a spectroscopic sample of galaxies at $0.03 \leq z \leq 0.11$ brighter than $M_{B} = -18.7$. The PM2GC sample comes from the Millennium Galaxy Catalogue \citep{Liske2003,Driver2005}, a deep $B$-band equatorial survey, complete down to $B = 20$ representative of the general field population in the local Universe. Utilising PM2GC data has the benefit that both the imaging instrumentation [Wide Field Camera (WFC) at Isaac Newton Telescope (INT)] and the technique used to determine the morphological types [MORPHOT] are identical for PM2GC and WINGS galaxies. This ensures that the two samples are fully morphologically consistent. We then computed the fraction of blue late-type galaxies brighter than $V=17.77$ in the field, which is $\sim 40$\,per cent, and we used this value to correct the cluster $N^{\rm BSp}$. 

Photometric RPS counts ($N_{\rm RPS}^{\rm BSp}$) were also corrected for field contamination. In the comparison field sample for the WINGS survey, PM2GC, only $6.7$\,per cent of the blue late-type with no gravitational interaction flags were classified as RPS candidates by \citetalias{Poggianti2016}. Poggianti (in prep) found that $86$\,per cent of the galaxies identified in clusters are genuine RPS. Therefore, we assume a similar success rate in this sample and consider $14$\,per cent of the candidates not genuine. A deduction of 0.14 is applied to the photometric fractions, to account for galaxy candidates that may not be undergoing ram-pressure stripping. We had a total of 39.6 blue late-type RPS candidates and 430.3 blue late-type galaxies, after correcting for the field contamination, in the 52 clusters of this sample (19 clusters had their photometric RPS fractions equal to zero). Equation~\ref{eq:phot_frac} defines the field-corrected photometric fractions of RPS candidate galaxies tabulated in Tab.~\ref{tab:cls_properties}%,     

\begin{equation}
\label{eq:phot_frac}
F_{\rm RPS}^{\rm phot}= \frac {N_{\rm RPS}^{\rm BSp} (1 - 0.067 - 0.14 )} {N^{\rm BSp} (1 - 0.4)}
\end{equation}

\subsection{Spectroscopic RPS fractions}
\label{sec:spec_frac}

To check if our photometric analysis was leading us to a false result, we also computed the spectroscopic RPS candidates fraction. In this analysis, we consider only confirmed spectroscopic members and clusters with spectroscopic completeness higher than $50$\,per cent. We have a total of 47 clusters in the spectroscopic sample.

The spectroscopic RPS fractions $F_{\rm RPS}^{\rm spec}$ were calculated by dividing the number of blue late-type RPS member candidates $M_{\rm RPS}^{\rm BSp}$ by the total number of blue late-type members $M^{\rm BSp}$ in that same aperture of $0.7\,R_{200}$, as described in Eq.~\ref{eq:spec_frac}. Overall we had 364 blue late-type members and 30.1 blue late-type RPS candidates (GASP survey success rate-corrected) tabulated in Tab.~\ref{tab:cls_properties}:

\begin{equation}
\label{eq:spec_frac}
F_{\rm RPS}^{\rm spec} = \frac {M_{\rm RPS}^{\rm BSp} (1 - 0.14) } {M^{\rm BSp}}
\end{equation}

This method is similar to the method applied in \citetalias{Vulcani2022}, with the exception that we restricted the physical radius and only used spectroscopically complete clusters in our analysis. 

\begin{figure}
	\includegraphics[width=\columnwidth]{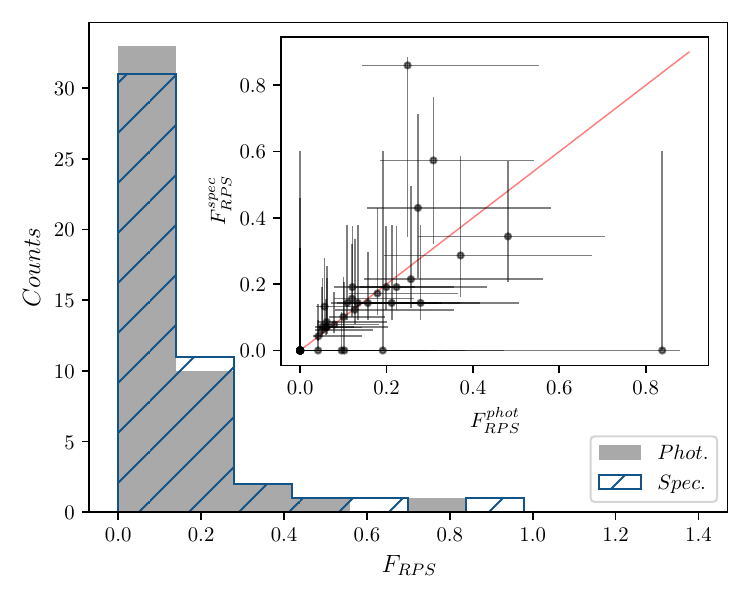}
    \caption{Comparison between photometric and spectroscopic RPS candidates fraction distributions. The grey histogram shows the distribution of the photometric RPS candidate fractions, while the blue hashed histogram shows the distribution of the spectroscopic RPS candidate fractions. The subplot in the top right corner compares photometric and spectroscopic RPS fractions for the common clusters in both samples. The 1:1 reference line is shown in red, and the binomial uncertainties are shown as black error bars.}
    \label{fig:comparison_frac}
\end{figure}

In Fig.~\ref{fig:comparison_frac}, we compare the distributions of photometric and spectroscopic RPS candidates fractions. Both distributions reach similar values and are comparable. In fact, the mean RPS candidate fraction of the photometric sample ($12 \pm 2$\,per cent) is consistent with the spectroscopic one ($11 \pm 2$\,per cent).
We computed the binomial confidence intervals following \citet{Cameron2011}. The method employed for calculating our confidence intervals is specifically designed for smaller fractions and yields reliable results for sample sizes ranging from small to intermediate. All uncertainties reported in this paper will be presented at a confidence level of c = 0.683, which is equivalent to $1\sigma$. Despite a few outliers, we found that $F_{\rm RPS}^{\rm spec}$ fall close to a $1:1$ ratio with the photometric fractions. The distributions of the photometric and spectroscopic RPS fractions are very similar (see Fig.~\ref{fig:comparison_frac}).

\begin{figure*}
	\includegraphics[width=17.8cm]{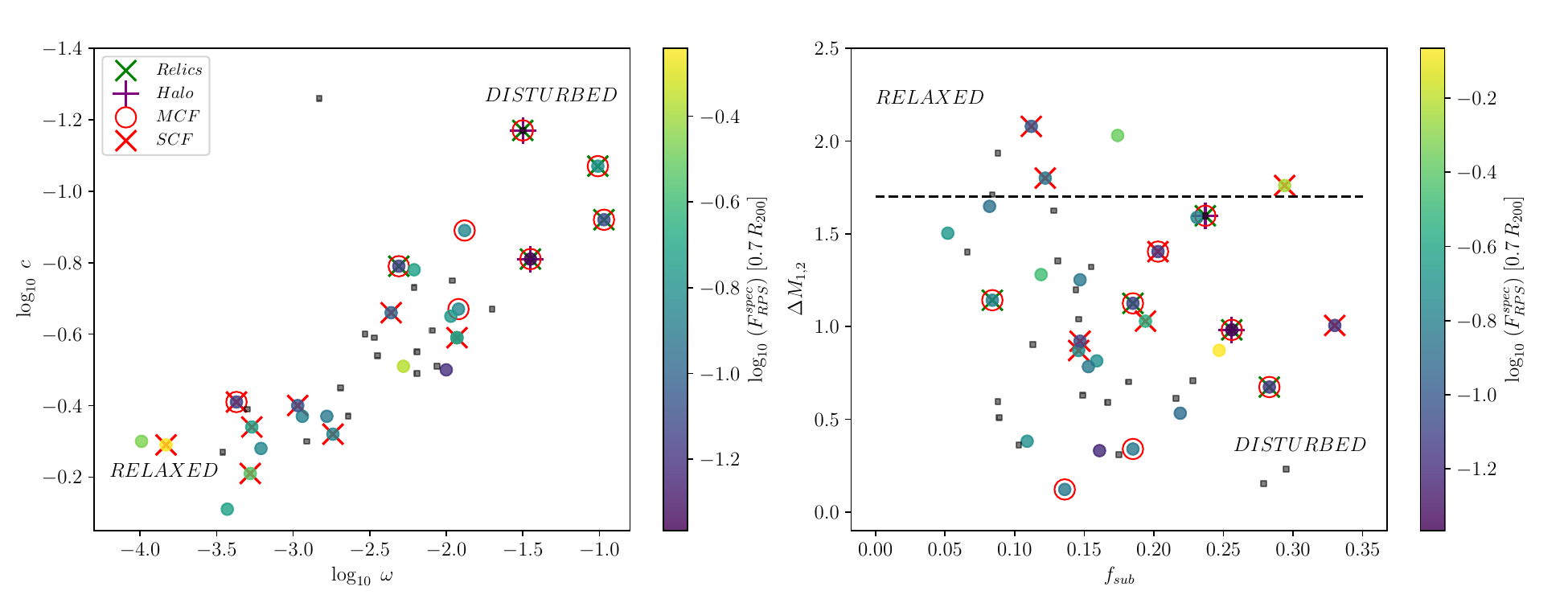}
    \caption{The same dynamical state diagnostics as in Fig.~\ref{fig:Yuan_flow_chart}, this time coloured according to the logarithm of the \textbf{spectroscopic} fractions of galaxies undergoing ram-pressure stripping. Clusters with $F_{\rm RPS}^{\rm spec} = 0$ are shown as black squares. Only clusters that are spectroscopically complete are plotted. The highest RPS candidate fractions very often lie on the more relaxed quadrant.}
    \label{fig:spec_frac}
\end{figure*}

Our main goal was to see how the RPS fractions compared with cluster dynamical state, so we started by comparing the $F_{\rm RPS}^{\rm spec}$ with the eight different proxies for cluster dynamical state described in Sec.~\ref{fig:cl_dyn_stage} and obtained no clear correlation, as can be seen in Fig.~\ref{fig:cl_dyn_stage} through the low correlation coefficients given between $F_{\rm RPS}^{\rm spec}$ and the other variables. 
Fig.~\ref{fig:spec_frac} shows this result in more detail, with X-ray (left) and optical (right) dynamical state diagnostics similar to Fig.~\ref{fig:Yuan_flow_chart}, but this time colour-coded by $F_{\rm RPS}^{\rm spec}$. We use spectroscopic fractions because they are more reliable than photometric ones, as they do not need statistical correction for the field contamination. The lack of correlation between the $F_{\rm RPS}^{\rm spec}$ fractions and the dynamical state of the cluster is still visible. We can even see that the highest RPS fractions sometimes appear in the most relaxed quadrant (bottom left) of the scatter plot on the left panel. We note that the results shown in Fig.~\ref{fig:spec_frac} and Fig.~\ref{fig:cl_dyn_stage} do not significantly change when using $F_{\rm RPS}^{\rm phot}$ instead of $F_{\rm RPS}^{\rm spec}$.

\begin{figure}
	\includegraphics[width=\columnwidth]{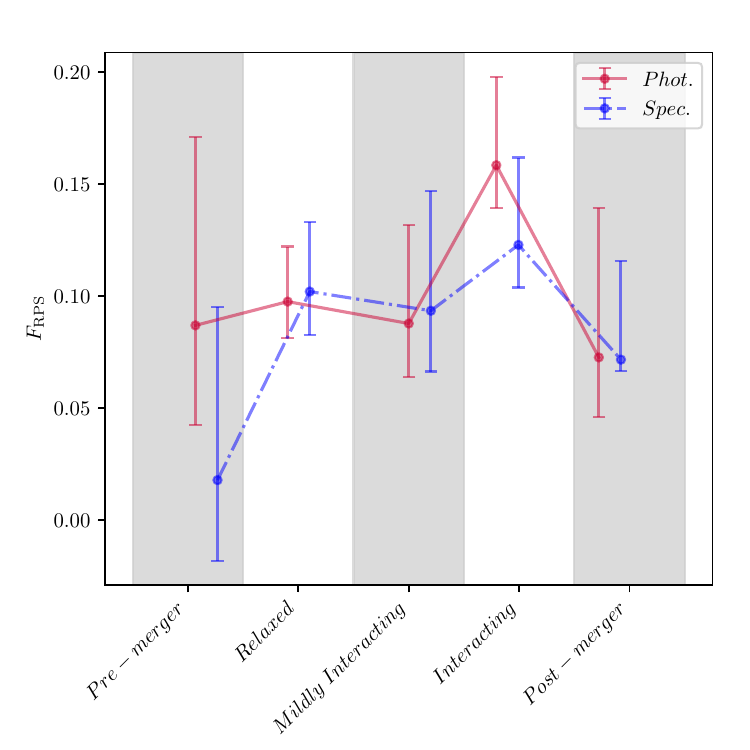}
    \caption{Fraction of RPS candidates relative to the infalling population as a function of the dynamical state sequence defined visually (see flowchart). The dashed blue and the solid red lines show the spectroscopic and photometric RPS candidate fractions, respectively. The $1\sigma$ binomial confidence intervals are shown. There is a possible enhancement in the RPS fractions in the interacting class, but it is not statistically significant.}
    \label{fig:comparison_frac_fl2}
\end{figure}

In Fig.~\ref{fig:comparison_frac_fl2}, we further study how the (photometric and spectroscopic) RPS fractions vary along the visually-defined dynamical state sequence. Although the spectroscopically complete cluster sample is smaller than the one used for the photometric analysis, the RPS fraction distributions are similar. Interestingly, when using the visual classification for the dynamical state in a sequence, we find a constant RPS candidate fraction from the "pre-merger" state to the "relaxed" and "mildly interacting" ones, and hints of an increase of the RPS candidate fraction for "interacting" clusters, followed by an equally mild decrease of the fraction for the "post-merger" class. The possible RPS fraction enhancement in the interacting class and subsequent decay in the post-merger class could be interpreted as a consequence of either the rapid timescales involved in the stripping process in a post-merging environment or the searched area not covering the radius where the bulk of RPS happens in these violent environments. Unfortunately, the sample is too small to allow drawing any significant conclusion from it. A Kruskal-Wallis H test \citep{Kruskal1952} showed that there was no statistically significant difference in RPS fractions between the different dynamical state classes.

%%%%%%%%%%%%%%%%%%%%%%%%%%%%%%%%%%%%%%%%%%%%%

\section{Discussion}
\label{sec:discussion}

In this paper, we have studied the influence of cluster buildup on the quenching of galaxies via RPS, using, for the first time, a sizeable homogeneous sample of stripped galaxies in a large sample of clusters. In particular, we computed the incidence of RPS candidates relative to the infalling blue late-type population within a fixed physical radius for 52 clusters with different dynamical states, determined using a variety of X-ray and optical proxies. We found no clear correlation with any of the dynamical state proxies used. The lack of correlation is seen when using photometric data (and correcting for field contamination) as well as when using confirmed spectroscopic members only. However, after visually classifying the sample in a dynamical state sequence, we observe a constant fraction going from "pre-mergers" to "relaxed" and "mildly interacting" clusters, and a possible (mild) enhancement in the RPS candidate fractions of the interacting clusters, which then goes back to the nominal value for the "post-merger" class (see Fig.~\ref{fig:comparison_frac_fl2}). The results are not statistically significant, but if they were to be confirmed, they point to an enhancement in the RPS (post-processing) in ongoing cluster mergers which occur on a fast timescale, but there are many caveats involved, discussed below.

There could be several explanations for the mild RPS increase in the interacting sample with respect to the other classes, in particular, the post-merger bin, which are associated with the many challenges in the study of galaxy populations in merging clusters, listed and discussed in the following: 

\begin{itemize}[wide,labelwidth=!,itemindent=!,labelindent=0pt, leftmargin=0em]

\item \textit{Cluster coverage:} In significantly disturbed clusters, mergers or post-mergers, much of the action (including RPS) could be happening outside the core of the cluster ($r > 0.7\,R_{200}$), where we do not have homogeneous data coverage. Only in clusters that are merging along the line of sight (LOS) we could see the merger close to the core. 
    
    LOS mergers are generally better traced by the optical diagnostics, such as $f_{\rm sub}$ and $\Delta M_{1,2}$. The fact that we do not find a clear correlation between the cluster dynamical state and the RPS candidates incidence using this kind of diagnostics plays against this hypothesis. 
    
    \item \textit{Cluster size:}
    To compare the incidence of any galaxy population between clusters, we need to search around the same physical radius (e.g $R_{200}$). However, in disturbed clusters, the physical size measurement can be unreliable (and has a different meaning), as its often derived from the dynamical mass (i.e. velocity dispersion). A more reliable alternative to estimate cluster mass in these cases is through gravitational lensing, which is beyond the scope of this manuscript.

    \item \textit{Cluster mass:} 
    Before drawing conclusions about the influence of the cluster dynamical states in the incidence of RPS, it is essential to check for possible correlations with cluster mass. RPS depends on the density of the ICM and is quadratically dependent on the velocity of the galaxies relative to the ICM, and both those quantities are higher in massive clusters, so one can naively expect a higher incidence of RPS galaxies in more massive clusters. However, there are contradictory results in the literature. 
    
    In recent work, \citet{Roberts2021a, Roberts2021b} identified jellyfish galaxies in $\sim$500 low-mass galaxy groups and 29 clusters ($M_{180}$ $\sim$ $10^{12.5}$ to $\sim$ $10^{15}$\,$M_{\odot}$) using radio continuum data. They found a factor of 2 increase in the fraction of stripped galaxies with respect to the star-forming galaxy population from groups to clusters but with significant scatter within the cluster sample.

    \citet{Wang2020} used HI-detected galaxies from the Arecibo Legacy Fast ALFA (ALFALFA) survey \citep{Haynes2018} in 26 X-ray-selected clusters ($M_{200}$ $\sim$ 3.8 $\times$ $10^{13}$ to $\sim$ 1.3 $\times$ $10^{15}$\,$M_{\odot}$) to estimate the ram-pressure strength at different radii of their cluster sample. They found that the RPS strength depends on the cluster mass and the projected phase space diagram position. They also computed the fractions of HI-rich galaxies affected by RPS. Their RPS fractions are higher in the massive clusters ($M_{200}$ > 2 $\times$ $10^{14}$\,$M_{\odot}$). They also showed that galaxies could undergo RPS at larger radii in massive clusters, in agreement with other studies \citep[e.g.][]{Jaffe2018,Gullieuszik2020,Pallero2019,Pallero2022}. 
    
    In the optical, \citetalias{Vulcani2022} found no correlation between the fraction of RPS galaxies relative to the infalling population of blue spirals and the cluster velocity dispersion, $\sigma_{\rm cl}$, or the cluster X-ray luminosity, $L_{\rm X}$. This could, in principle, be affected by interacting clusters with unreliable masses. To account for this, in Fig.~\ref{fig:mass_control2}, we investigated whether the spectroscopic fractions of the RPS candidates we have computed show any dependence on the halo mass and X-ray luminosity measured in the $0.1$ -- $2.4$\,keV energy band for \textit{ROSAT} data in \citet{Ebeling_1996,Ebeling_1998,Ebeling_2000} excluding interacting clusters. In fact, we only consider the relaxed and mildly-interacting clusters from our flowchart dynamical state sequence. Even when cleaning the sample, we still do not find a correlation between the incidence of RPS candidates and cluster mass. 
    
    In summary, our optical quantification of the incidence of RPS galaxies in clusters does not show the same trends with cluster mass as studies in radio wavelengths. Low-frequency radio continuum can be more sensitive to RPS than the optical, as the non-thermal interstellar medium components appear more affected by RPS \citep{Roberts2021a,Roberts2021b,Ignesti_2022,Ignesti2023}. Another explanation for the discrepancy between the wavelengths is that optical selection favours galaxies with star formation in the tail. Our sample selection might underestimate the total number of galaxies. Some galaxies at peak stripping show a lack of star formation in the tail \citep[e.g.][]{Boselli_2016,Laudari_2022}. Additionally, optical wavelengths only allow us to observe galaxies that had time to form new stars from the stripped gas \citep[e.g.][]{Poggianti_2019,Gullieuszik2020}. Our results are in line with \citet{Gullieuszik2020}, who used MUSE data from the GASP survey to analyse the dependency of the SFR in the tails of stripped gas on galaxy and clusters' properties, including cluster mass. They conclude that the interplay between all the parameters involved is complex and that there is no single dominant parameter impacting the observed SFR amount. Their conclusion could be related to our findings on the global RPS candidate fractions.

\begin{figure}
	\includegraphics[width=\columnwidth]{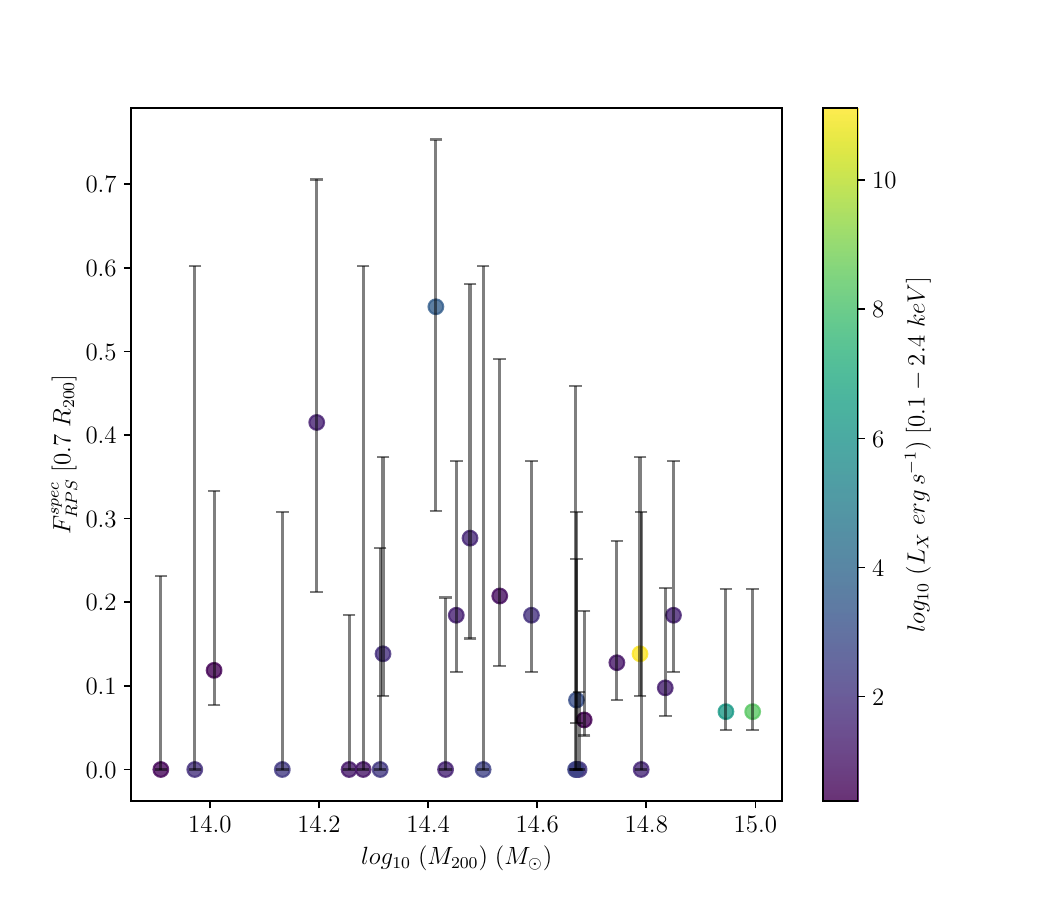}
    \caption{Cluster RPS spectroscopic fractions vs. halo masses coloured by the cluster X-ray luminosity measured in the $0.1$ -- $2.4$\,keV energy band for \textit{ROSAT} data in \citet{Ebeling_1996,Ebeling_1998,Ebeling_2000}. In this plot, we are showing only the relaxed and mildly-interacting clusters. Massive clusters are not the ones with the highest RPS candidate fractions. The $1\sigma$ confidence interval is computed as in \citet{Cameron2011}}
    \label{fig:mass_control2}
\end{figure}

    \item \textit{Cluster members:} Regarding the previous point, the disturbed clusters also challenge defining true members of very close pre-mergers. We dealt with this by treating systems that are separated by less than $R_{200}$ as if they were the same, but we expect some contamination from interlopers. 

    \item \textit{Cluster centres:} In relaxed clusters, the X-ray peak, centroids, BCG, and density peak are close to each other. However, during merging events, the galaxy density peak can be very displaced from the X-ray centroid \citep[$\sim$ 0.2\,Mpc for the bullet cluster,][]{Clowe_2006}. In this work, we used the symmetry of the cluster system to determine its centre. We assumed that the midpoint between the two central brightest galaxies in a cluster system represents the system's centre of mass better than the X-ray peak or the BCG. 

    \item \textit{Galaxy populations:} Using HST images for a few intermediate-z clusters, previous studies such as \citet{RomanOliveira2019,Owers2012,Ebeling2019,Rawle2014} have focused on the absolute number or RPS features strength of stripped galaxies without accounting for the size of this population relative to the infalling blue late-type galaxies or the physical size of the explored region. On the contrary, in our study, we compute for the first time in a homogeneous cluster sample the fraction of galaxies undergoing RPS relative to the infalling population in a fixed aperture, normalised by cluster size. \citetalias{Vulcani2022} estimated similar RPS candidate fractions using the  \citetalias{Poggianti2016} sample and additional candidates in the same clusters. The main difference with our study is that they did not take into account the cluster size or limited their sample to clusters with spectroscopic completeness $> 50$\,per cent, as we did in this work. Also, we only considered \citetalias{Poggianti2016} galaxies and not the new candidates since these were searched only among spectroscopic members, and we used photometric and spectroscopic fractions, which would have been difficult to compare otherwise. Additionally, part of their new RPS candidates are unwinding spirals. When the \citetalias{Poggianti2016} sample was compiled, the unwinding effect of ram-pressure stripping on the spiral arms of galaxies was still unknown \citep{Bellhouse2021}. \citetalias{Vulcani2022} have found that about $35$\,per cent of the total population of non-interacting blue late-type cluster galaxies are undergoing RPS. When using only the \citetalias{Poggianti2016} candidates they obtain $\sim$ $15-20$\,per cent. Despite the different methods, this is close to the spectroscopic and photometric RPS fractions we measure within $0.7 R_{200}$, which are $\sim$ $11-12$\,per cent. Given that \citetalias{Vulcani2022} found new RPS candidates in the same clusters, our fractions can be considered lower limits.

\end{itemize}

\section{Summary and Conclusions}
\label{sec:conclusions}

This work studies the incidence of RPS candidates relative to the infalling population of galaxies in clusters of different dynamical states using the largest and most homogeneous sample of RPS candidates at low redshift from \citetalias{Poggianti2016} and a variety of cluster dynamical state proxies. 

In summary, the main results are:
\begin{itemize}[wide,labelwidth=!,itemindent=!,labelindent=0pt, leftmargin=0em]

\item We computed the photometric and spectroscopic fractions of RPS candidates relative to the infalling population of blue spiral galaxies in each cluster. The former was carefully corrected by the expected field contamination, while the latter used confirmed cluster members. They both considered the same circular area within the clusters (with a radius of $0.7\,R_{200}$) for homogeneity. Both fractions are in general agreement with some scatter. The mean RPS candidate fraction of both samples is around $11-12$\,per cent.

\item We found no correlation between RPS candidate fractions and a variety of automatic cluster dynamical state indicators. In particular, we examined eight optical and X-ray indicators, including X-ray concentration and the magnitude gap between the two brightest galaxies. 

\item We further constructed a "cluster merger sequence" by splitting the sample into five different dynamical state classes (pre-merger, relaxed, interacting, interacting and post-merger) following a visual dynamical state classification flowchart based on optical, X-ray, and radio data. When doing this, we found hints of an RPS candidate fraction enhancement in the interacting class, followed by a decay in the post-merger bin. The results, however, are driven by a limited number of interacting and post-merger clusters (five and five) and are not statistically significant.

\item We found no apparent correlation between RPS candidate fractions and cluster mass or X-ray luminosity, even when excluding the disturbed clusters from our sample. This could be associated with the small radius coverage of our analysis or the complexity of the RPS process. 
\end{itemize}   

A few studies of individual interacting clusters found elevated numbers of galaxies' RPS features. These studies, together with theoretical predictions, built a common understanding that collisions between clusters would increase the fractions of galaxies undergoing RPS. This work is the first attempt to quantify the effect of cluster dynamical stage in RPS in a homogeneous way using a large cluster sample. Overall, we find hints of a possible RPS enhancement in interacting clusters. %, which decays in post-merging systems. 
The low confidence in our results is due to the inevitably low-number statistics (despite using the largest homogeneous sample available) and the complexity of cluster growth. In particular, the limited coverage of cluster observations and the nature of the optical selection of RPS galaxies, which requires star formation in the striped tails.

Our study, with its methods and special considerations, can serve as a guide for future studies using even larger samples of RPS candidates currently being assembled\footnote{See, for example, the citizen science project "Fishing for jellyfish galaxies": \url{https://www.zooniverse.org/projects/cbellhouse/fishing-for-jellyfish-galaxies}}, and also wider multi-wavelength cluster coverage, such as the upcoming WEAVE \citep[][]{Jin2023} and CHANCES \citep{Haines2023} wide-field cluster spectroscopic surveys and the eROSITA  X-ray mission \citep{Merloni2012}.

\section*{Acknowledgements}

We thank the anonymous referee for their valuable comments.
ACCL thanks the financial support of the National Agency for Research and Development (ANID) / Scholarship Program / DOCTORADO BECAS CHILE/2019-21190049. ACCL thanks Rosa Calvi for guiding us in obtaining data from the PM2GC catalogue. ACCL thanks Elke R\"odiger and Diego Pallero for fruitful conversations about this work. 
YJ and ACCL acknowledge financial support from ANID BASAL project No. FB210003 and FONDECYT Iniciaci\'on 2018 No. 11180558. 
BV and MG acknowledge financial contribution from the grant PRIN MIUR 2017 n.20173ML3WW\_001 (PI Cimatti), from the INAF main-stream funding programme (PI B. Vulcani). We acknowledge funding from the INAF main-stream funding programme (PI B. Vulcani). This project has received funding from the European Research Council (ERC) under the Horizon 2020 research and innovation programme (grant agreement N. 833824). 
We acknowledge the financial contribution from the agreement ASI-INAF n.2017-14-H.0 (PI Moretti). 
KK acknowledges full financial support from ANID through FONDECYT Postdoctrorado Project 3200139. 
JPC acknowledges financial support from ANID through FONDECYT Postdoctorado Project 3210709.
AI acknowledges the INAF founding program 'Ricerca Fondamentale 2022' (PI A. Ignesti).
SLM acknowledges support from the Science and Technology Facilities Council through grant number ST/N021702/1. 
AW acknowledges financial support from ASI through the ASI-INAF agreements 2017-14-H.0.

This research has made use of "Aladin sky atlas" developed at CDS, Strasbourg Observatory, France \citep{2000A&AS..143...33B, 2014ASPC..485..277B}. This research has made use of the VizieR catalogue access tool, CDS, Strasbourg, France. This research made use of TOPCAT, an interactive graphical viewer and editor for tabular data \citep{2005ASPC..347...29T}. This research made use of ds9, a tool for data visualisation supported by the {\it Chandra} X-ray Science Center (CXC) and the High Energy Astrophysics Science Archive Center (HEASARC) with support from the JWST Mission office at the Space Telescope Science Institute for 3D visualisation. This research made use of Astropy,\footnote{http://www.astropy.org} a community-developed core Python package for Astronomy \citep{astropy:2013, astropy:2018}.

%%%%%%%%%%%%%%%%%%%%%%%%%%%%%%%%%%%%%%%%%%%%%%%%%%
\section*{Data Availability}

The data underlying this manuscript can be retrieved in Vizier \url{https://vizier.cds.unistra.fr/viz-bin/VizieR}, searching for the words WINGS, OmegaWINGS to retrieve the photometric and spectroscopic catalogues. PM2GC, and jellyfish in the find catalogue field. Any extra data product will be shared upon reasonable request to the corresponding author.



\bibliographystyle{mnras}
\bibliography{refs} % if your bibtex file is called example.bib

\begin{thebibliography}{}
\makeatletter
\relax
\def\mn@urlcharsother{\let\do\@makeother \do\$\do\&\do\#\do\^\do\_\do\%\do\~}
\def\mn@doi{\begingroup\mn@urlcharsother \@ifnextchar [ {\mn@doi@}
  {\mn@doi@[]}}
\def\mn@doi@[#1]#2{\def\@tempa{#1}\ifx\@tempa\@empty \href
  {http://dx.doi.org/#2} {doi:#2}\else \href {http://dx.doi.org/#2} {#1}\fi
  \endgroup}
\def\mn@eprint#1#2{\mn@eprint@#1:#2::\@nil}
\def\mn@eprint@arXiv#1{\href {http://arxiv.org/abs/#1} {{\tt arXiv:#1}}}
\def\mn@eprint@dblp#1{\href {http://dblp.uni-trier.de/rec/bibtex/#1.xml}
  {dblp:#1}}
\def\mn@eprint@#1:#2:#3:#4\@nil{\def\@tempa {#1}\def\@tempb {#2}\def\@tempc
  {#3}\ifx \@tempc \@empty \let \@tempc \@tempb \let \@tempb \@tempa \fi \ifx
  \@tempb \@empty \def\@tempb {arXiv}\fi \@ifundefined
  {mn@eprint@\@tempb}{\@tempb:\@tempc}{\expandafter \expandafter \csname
  mn@eprint@\@tempb\endcsname \expandafter{\@tempc}}}

\bibitem[\protect\citeauthoryear{{Akamatsu}, {Takizawa}, {Nakazawa},
  {Fukazawa}, {Ishisaki}  \& {Ohashi}}{{Akamatsu} et~al.}{2012}]{Akamatsu2012}
{Akamatsu} H.,  {Takizawa} M.,  {Nakazawa} K.,  {Fukazawa} Y.,  {Ishisaki} Y.,
   {Ohashi} T.,  2012, \mn@doi [\pasj] {10.1093/pasj/64.4.67}, \href
  {http://adsabs.harvard.edu/abs/2012PASJ...64...67A} {64, 67}

\bibitem[\protect\citeauthoryear{{Andrade-Santos} et~al.,}{{Andrade-Santos}
  et~al.}{2015}]{Andrade2015}
{Andrade-Santos} F.,  et~al., 2015, \mn@doi [\apj]
  {10.1088/0004-637X/803/2/108}, \href
  {https://ui.adsabs.harvard.edu/abs/2015ApJ...803..108A} {803, 108}

\bibitem[\protect\citeauthoryear{{Andrade-Santos} et~al.,}{{Andrade-Santos}
  et~al.}{2017}]{Andrade_Santos_2017}
{Andrade-Santos} F.,  et~al., 2017, \mn@doi [\apj] {10.3847/1538-4357/aa7461},
  \href {https://ui.adsabs.harvard.edu/abs/2017ApJ...843...76A} {843, 76}

\bibitem[\protect\citeauthoryear{{Ascasibar} \& {Markevitch}}{{Ascasibar} \&
  {Markevitch}}{2006}]{Ascasibar2006}
{Ascasibar} Y.,  {Markevitch} M.,  2006, \mn@doi [\apj] {10.1086/506508}, \href
  {https://ui.adsabs.harvard.edu/abs/2006ApJ...650..102A} {650, 102}

\bibitem[\protect\citeauthoryear{{Astropy Collaboration} et~al.,}{{Astropy
  Collaboration} et~al.}{2013}]{astropy:2013}
{Astropy Collaboration} et~al., 2013, \mn@doi [\aap]
  {10.1051/0004-6361/201322068}, \href
  {http://adsabs.harvard.edu/abs/2013A%26A...558A..33A} {558, A33}

\bibitem[\protect\citeauthoryear{{Astropy Collaboration} et~al.,}{{Astropy
  Collaboration} et~al.}{2018}]{astropy:2018}
{Astropy Collaboration} et~al., 2018, \mn@doi [\aj] {10.3847/1538-3881/aabc4f},
  \href {https://ui.adsabs.harvard.edu/abs/2018AJ....156..123A} {156, 123}

\bibitem[\protect\citeauthoryear{{Bacchi}, {Feretti}, {Giovannini}  \&
  {Govoni}}{{Bacchi} et~al.}{2003}]{Bacchi2003}
{Bacchi} M.,  {Feretti} L.,  {Giovannini} G.,   {Govoni} F.,  2003, \mn@doi
  [\aap] {10.1051/0004-6361:20030044}, \href
  {https://ui.adsabs.harvard.edu/abs/2003A&A...400..465B} {400, 465}

\bibitem[\protect\citeauthoryear{{Bagchi}, {Durret}, {Neto}  \&
  {Paul}}{{Bagchi} et~al.}{2006}]{Bagchi2006}
{Bagchi} J.,  {Durret} F.,  {Neto} G. B.~L.,   {Paul} S.,  2006, \mn@doi
  [Science] {10.1126/science.1131189}, \href
  {https://ui.adsabs.harvard.edu/abs/2006Sci...314..791B} {314, 791}

\bibitem[\protect\citeauthoryear{{Baldi}, {Bardelli}  \& {Zucca}}{{Baldi}
  et~al.}{2001}]{Baldi2001}
{Baldi} A.,  {Bardelli} S.,   {Zucca} E.,  2001, \mn@doi [\mnras]
  {10.1046/j.1365-8711.2001.04358.x}, \href
  {https://ui.adsabs.harvard.edu/abs/2001MNRAS.324..509B} {324, 509}

\bibitem[\protect\citeauthoryear{{Bardelli}, {Zucca}  \& {Baldi}}{{Bardelli}
  et~al.}{2001}]{Bardelli2001}
{Bardelli} S.,  {Zucca} E.,   {Baldi} A.,  2001, \mn@doi [\mnras]
  {10.1046/j.1365-8711.2001.03973.x}, \href
  {https://ui.adsabs.harvard.edu/abs/2001MNRAS.320..387B} {320, 387}

\bibitem[\protect\citeauthoryear{{Bekki}}{{Bekki}}{2009}]{Bekki2009}
{Bekki} K.,  2009, \mn@doi [\mnras] {10.1111/j.1365-2966.2009.15431.x}, \href
  {https://ui.adsabs.harvard.edu/abs/2009MNRAS.399.2221B} {399, 2221}

\bibitem[\protect\citeauthoryear{{Bellhouse} et~al.,}{{Bellhouse}
  et~al.}{2021}]{Bellhouse2021}
{Bellhouse} C.,  et~al., 2021, \mn@doi [\mnras] {10.1093/mnras/staa3298}, \href
  {https://ui.adsabs.harvard.edu/abs/2021MNRAS.500.1285B} {500, 1285}

\bibitem[\protect\citeauthoryear{Benavides, Sales  \& Abadi}{Benavides
  et~al.}{2020}]{Benavides2020}
Benavides J.~A.,  Sales L.~V.,   Abadi M.~G.,  2020, \mn@doi [Monthly Notices
  of the Royal Astronomical Society] {10.1093/mnras/staa2636}, 498, 3852

\bibitem[\protect\citeauthoryear{{Benavides}, {Biviano}  \&
  {Abadi}}{{Benavides} et~al.}{2023}]{Benavides2023}
{Benavides} J.~A.,  {Biviano} A.,   {Abadi} M.~G.,  2023, \mn@doi [\aap]
  {10.1051/0004-6361/202245422}, \href
  {https://ui.adsabs.harvard.edu/abs/2023A&A...669A.147B} {669, A147}

\bibitem[\protect\citeauthoryear{{Bilton}, {Hunt}, {Pimbblet}  \&
  {Roediger}}{{Bilton} et~al.}{2019}]{Bilton2019}
{Bilton} L.~E.,  {Hunt} M.,  {Pimbblet} K.~A.,   {Roediger} E.,  2019, \mn@doi
  [\mnras] {10.1093/mnras/stz2927}, \href
  {https://ui.adsabs.harvard.edu/abs/2019MNRAS.490.5017B} {490, 5017}

\bibitem[\protect\citeauthoryear{{Birnboim}, {Keshet}  \&
  {Hernquist}}{{Birnboim} et~al.}{2010}]{Birnboim2010}
{Birnboim} Y.,  {Keshet} U.,   {Hernquist} L.,  2010, \mn@doi [\mnras]
  {10.1111/j.1365-2966.2010.17176.x}, \href
  {https://ui.adsabs.harvard.edu/abs/2010MNRAS.408..199B} {408, 199}

\bibitem[\protect\citeauthoryear{{Biviano} et~al.,}{{Biviano}
  et~al.}{2017}]{Biviano_2017}
{Biviano} A.,  et~al., 2017, \mn@doi [\aap] {10.1051/0004-6361/201731289},
  \href {https://ui.adsabs.harvard.edu/abs/2017A&A...607A..81B} {607, A81}

\bibitem[\protect\citeauthoryear{{Boch} \& {Fernique}}{{Boch} \&
  {Fernique}}{2014}]{2014ASPC..485..277B}
{Boch} T.,  {Fernique} P.,  2014, in {Manset} N.,  {Forshay} P.,  eds,
  Astronomical Society of the Pacific Conference Series, Vol.~485, Astronomical
  Data Analysis Software and Systems XXIII.
ASP, p.~277

\bibitem[\protect\citeauthoryear{{Bonnarel} et~al.,}{{Bonnarel}
  et~al.}{2000}]{2000A&AS..143...33B}
{Bonnarel} F.,  et~al., 2000, \mn@doi [\aaps] {10.1051/aas:2000331}, \href
  {https://ui.adsabs.harvard.edu/abs/2000A&AS..143...33B} {143, 33}

\bibitem[\protect\citeauthoryear{{Boselli} \& {Gavazzi}}{{Boselli} \&
  {Gavazzi}}{2006}]{Boselli2006}
{Boselli} A.,  {Gavazzi} G.,  2006, \mn@doi [\pasp] {10.1086/500691}, \href
  {https://ui.adsabs.harvard.edu/abs/2006PASP..118..517B} {118, 517}

\bibitem[\protect\citeauthoryear{{Boselli} et~al.,}{{Boselli}
  et~al.}{2016}]{Boselli_2016}
{Boselli} A.,  et~al., 2016, \mn@doi [\aap] {10.1051/0004-6361/201527795},
  \href {https://ui.adsabs.harvard.edu/abs/2016A&A...587A..68B} {587, A68}

\bibitem[\protect\citeauthoryear{{Boselli}, {Fossati}  \& {Sun}}{{Boselli}
  et~al.}{2021}]{Boselli2021}
{Boselli} A.,  {Fossati} M.,   {Sun} M.,  2021, arXiv e-prints, \href
  {https://ui.adsabs.harvard.edu/abs/2021arXiv210913614B} {p. arXiv:2109.13614}

\bibitem[\protect\citeauthoryear{{Botteon}, {Gastaldello}  \&
  {Brunetti}}{{Botteon} et~al.}{2018}]{Botteon2018}
{Botteon} A.,  {Gastaldello} F.,   {Brunetti} G.,  2018, \mn@doi [\mnras]
  {10.1093/mnras/sty598}, \href
  {https://ui.adsabs.harvard.edu/abs/2018MNRAS.476.5591B} {476, 5591}

\bibitem[\protect\citeauthoryear{{Briel}, {Finoguenov}  \& {Henry}}{{Briel}
  et~al.}{2004}]{Briel2004}
{Briel} U.~G.,  {Finoguenov} A.,   {Henry} J.~P.,  2004, \mn@doi [\aap]
  {10.1051/0004-6361:20035812}, \href
  {https://ui.adsabs.harvard.edu/abs/2004A&A...426....1B} {426, 1}

\bibitem[\protect\citeauthoryear{{Brunetti} \& {Jones}}{{Brunetti} \&
  {Jones}}{2014}]{Brunetti2014}
{Brunetti} G.,  {Jones} T.~W.,  2014, \mn@doi [International Journal of Modern
  Physics D] {10.1142/S0218271814300079}, \href
  {https://ui.adsabs.harvard.edu/abs/2014IJMPD..2330007B} {23, 1430007}

\bibitem[\protect\citeauthoryear{{Byrd} \& {Valtonen}}{{Byrd} \&
  {Valtonen}}{1990}]{Byrd1990}
{Byrd} G.,  {Valtonen} M.,  1990, \mn@doi [\apj] {10.1086/168362}, \href
  {https://ui.adsabs.harvard.edu/abs/1990ApJ...350...89B} {350, 89}

\bibitem[\protect\citeauthoryear{{Calvi}, {Poggianti}  \& {Vulcani}}{{Calvi}
  et~al.}{2011}]{Calvi2011}
{Calvi} R.,  {Poggianti} B.~M.,   {Vulcani} B.,  2011, \mn@doi [\mnras]
  {10.1111/j.1365-2966.2011.19088.x}, \href
  {https://ui.adsabs.harvard.edu/abs/2011MNRAS.416..727C} {416, 727}

\bibitem[\protect\citeauthoryear{{Calvi}, {Poggianti}, {Fasano}  \&
  {Vulcani}}{{Calvi} et~al.}{2012}]{Calvi2012}
{Calvi} R.,  {Poggianti} B.~M.,  {Fasano} G.,   {Vulcani} B.,  2012, \mn@doi
  [\mnras] {10.1111/j.1745-3933.2011.01168.x}, \href
  {https://ui.adsabs.harvard.edu/abs/2012MNRAS.419L..14C} {419, L14}

\bibitem[\protect\citeauthoryear{{Cameron}}{{Cameron}}{2011}]{Cameron2011}
{Cameron} E.,  2011, \mn@doi [\pasa] {10.1071/AS10046}, \href
  {https://ui.adsabs.harvard.edu/abs/2011PASA...28..128C} {28, 128}

\bibitem[\protect\citeauthoryear{{Campitiello} et~al.,}{{Campitiello}
  et~al.}{2022}]{Campitiello2022}
{Campitiello} M.~G.,  et~al., 2022, \mn@doi [\aap]
  {10.1051/0004-6361/202243470}, \href
  {https://ui.adsabs.harvard.edu/abs/2022A&A...665A.117C} {665, A117}

\bibitem[\protect\citeauthoryear{{Cava} et~al.,}{{Cava}
  et~al.}{2009}]{Cava_2009}
{Cava} A.,  et~al., 2009, \mn@doi [\aap] {10.1051/0004-6361:200810997}, \href
  {https://ui.adsabs.harvard.edu/abs/2009A&A...495..707C} {495, 707}

\bibitem[\protect\citeauthoryear{{Choque-Challapa}, {Smith}, {Candlish},
  {Peletier}  \& {Shin}}{{Choque-Challapa} et~al.}{2019}]{Choque2019}
{Choque-Challapa} N.,  {Smith} R.,  {Candlish} G.,  {Peletier} R.,   {Shin} J.,
   2019, \mn@doi [\mnras] {10.1093/mnras/stz2829}, \href
  {https://ui.adsabs.harvard.edu/abs/2019MNRAS.490.3654C} {490, 3654}

\bibitem[\protect\citeauthoryear{{Clowe}, {Brada{\v{c}}}, {Gonzalez},
  {Markevitch}, {Randall}, {Jones}  \& {Zaritsky}}{{Clowe}
  et~al.}{2006}]{Clowe_2006}
{Clowe} D.,  {Brada{\v{c}}} M.,  {Gonzalez} A.~H.,  {Markevitch} M.,  {Randall}
  S.~W.,  {Jones} C.,   {Zaritsky} D.,  2006, \mn@doi [\apjl] {10.1086/508162},
  \href {https://ui.adsabs.harvard.edu/abs/2006ApJ...648L.109C} {648, L109}

\bibitem[\protect\citeauthoryear{{Conselice}}{{Conselice}}{2003}]{Conselice2003}
{Conselice} C.~J.,  2003, \mn@doi [\apjs] {10.1086/375001}, \href
  {https://ui.adsabs.harvard.edu/abs/2003ApJS..147....1C} {147, 1}

\bibitem[\protect\citeauthoryear{{Contreras-Santos} et~al.,}{{Contreras-Santos}
  et~al.}{2022}]{Contreras-Santos2022}
{Contreras-Santos} A.,  et~al., 2022, \mn@doi [\mnras] {10.1093/mnras/stac275},
  \href {https://ui.adsabs.harvard.edu/abs/2022MNRAS.511.2897C} {511, 2897}

\bibitem[\protect\citeauthoryear{{Cowie} \& {Songaila}}{{Cowie} \&
  {Songaila}}{1977}]{Cowie1977}
{Cowie} L.~L.,  {Songaila} A.,  1977, \mn@doi [\nat] {10.1038/266501a0}, \href
  {https://ui.adsabs.harvard.edu/abs/1977Natur.266..501C} {266, 501}

\bibitem[\protect\citeauthoryear{{Crossett}, {Pimbblet}, {Jones}, {Brown}  \&
  {Stott}}{{Crossett} et~al.}{2017}]{Crossett2017}
{Crossett} J.~P.,  {Pimbblet} K.~A.,  {Jones} D.~H.,  {Brown} M. J.~I.,
  {Stott} J.~P.,  2017, \mn@doi [\mnras] {10.1093/mnras/stw2228}, \href
  {https://ui.adsabs.harvard.edu/abs/2017MNRAS.464..480C} {464, 480}

\bibitem[\protect\citeauthoryear{{Crossett} et~al.,}{{Crossett}
  et~al.}{2022}]{Crossett2022}
{Crossett} J.~P.,  et~al., 2022, \mn@doi [\aap] {10.1051/0004-6361/202142057},
  \href {https://ui.adsabs.harvard.edu/abs/2022A&A...663A...2C} {663, A2}

\bibitem[\protect\citeauthoryear{{Dariush}, {Khosroshahi}, {Ponman}, {Pearce},
  {Raychaudhury}  \& {Hartley}}{{Dariush} et~al.}{2007}]{Dariush2007}
{Dariush} A.,  {Khosroshahi} H.~G.,  {Ponman} T.~J.,  {Pearce} F.,
  {Raychaudhury} S.,   {Hartley} W.,  2007, \mn@doi [\mnras]
  {10.1111/j.1365-2966.2007.12385.x}, \href
  {https://ui.adsabs.harvard.edu/abs/2007MNRAS.382..433D} {382, 433}

\bibitem[\protect\citeauthoryear{{De Luca}, {De Petris}, {Yepes}, {Cui},
  {Knebe}  \& {Rasia}}{{De Luca} et~al.}{2021}]{DeLuca2021}
{De Luca} F.,  {De Petris} M.,  {Yepes} G.,  {Cui} W.,  {Knebe} A.,   {Rasia}
  E.,  2021, \mn@doi [\mnras] {10.1093/mnras/stab1073}, \href
  {https://ui.adsabs.harvard.edu/abs/2021MNRAS.504.5383D} {504, 5383}

\bibitem[\protect\citeauthoryear{{Domainko} et~al.,}{{Domainko}
  et~al.}{2006}]{Domainko_2006}
{Domainko} W.,  et~al., 2006, \mn@doi [\aap] {10.1051/0004-6361:20053921},
  \href {https://ui.adsabs.harvard.edu/abs/2006A&A...452..795D} {452, 795}

\bibitem[\protect\citeauthoryear{{Dressler}}{{Dressler}}{1980}]{Dressler1980}
{Dressler} A.,  1980, \mn@doi [\apj] {10.1086/157753}, \href
  {https://ui.adsabs.harvard.edu/abs/1980ApJ...236..351D} {236, 351}

\bibitem[\protect\citeauthoryear{{Dressler} \& {Shectman}}{{Dressler} \&
  {Shectman}}{1988}]{DresslerShectman1988}
{Dressler} A.,  {Shectman} S.~A.,  1988, \mn@doi [\aj] {10.1086/114694}, \href
  {https://ui.adsabs.harvard.edu/abs/1988AJ.....95..985D} {95, 985}

\bibitem[\protect\citeauthoryear{{Driver}, {Liske}, {Cross}, {De Propris}  \&
  {Allen}}{{Driver} et~al.}{2005}]{Driver2005}
{Driver} S.~P.,  {Liske} J.,  {Cross} N.~J.~G.,  {De Propris} R.,   {Allen}
  P.~D.,  2005, \mn@doi [\mnras] {10.1111/j.1365-2966.2005.08990.x}, \href
  {https://ui.adsabs.harvard.edu/abs/2005MNRAS.360...81D} {360, 81}

\bibitem[\protect\citeauthoryear{{Durret}, {Perrot}, {Lima Neto}, {Adami},
  {Bertin}  \& {Bagchi}}{{Durret} et~al.}{2013}]{Durret2013}
{Durret} F.,  {Perrot} C.,  {Lima Neto} G.~B.,  {Adami} C.,  {Bertin} E.,
  {Bagchi} J.,  2013, \mn@doi [\aap] {10.1051/0004-6361/201322082}, \href
  {https://ui.adsabs.harvard.edu/abs/2013A&A...560A..78D} {560, A78}

\bibitem[\protect\citeauthoryear{{Durret}, {Chiche}, {Lobo}  \&
  {Jauzac}}{{Durret} et~al.}{2021}]{Durret2021}
{Durret} F.,  {Chiche} S.,  {Lobo} C.,   {Jauzac} M.,  2021, \mn@doi [\aap]
  {10.1051/0004-6361/202039770}, \href
  {https://ui.adsabs.harvard.edu/abs/2021A&A...648A..63D} {648, A63}

\bibitem[\protect\citeauthoryear{{Dwarakanath}, {Parekh}, {Kale}  \&
  {George}}{{Dwarakanath} et~al.}{2018}]{Dwarakanath2018}
{Dwarakanath} K.~S.,  {Parekh} V.,  {Kale} R.,   {George} L.~T.,  2018, \mn@doi
  [\mnras] {10.1093/mnras/sty744}, \href
  {https://ui.adsabs.harvard.edu/abs/2018MNRAS.477..957D} {477, 957}

\bibitem[\protect\citeauthoryear{{Ebeling} \& {Kalita}}{{Ebeling} \&
  {Kalita}}{2019}]{Ebeling2019}
{Ebeling} H.,  {Kalita} B.~S.,  2019, \mn@doi [\apj]
  {10.3847/1538-4357/ab35d6}, \href
  {https://ui.adsabs.harvard.edu/abs/2019ApJ...882..127E} {882, 127}

\bibitem[\protect\citeauthoryear{{Ebeling}, {Voges}, {Bohringer}, {Edge},
  {Huchra}  \& {Briel}}{{Ebeling} et~al.}{1996}]{Ebeling_1996}
{Ebeling} H.,  {Voges} W.,  {Bohringer} H.,  {Edge} A.~C.,  {Huchra} J.~P.,
  {Briel} U.~G.,  1996, \mn@doi [\mnras] {10.1093/mnras/281.3.799}, \href
  {https://ui.adsabs.harvard.edu/abs/1996MNRAS.281..799E} {281, 799}

\bibitem[\protect\citeauthoryear{{Ebeling}, {Edge}, {Bohringer}, {Allen},
  {Crawford}, {Fabian}, {Voges}  \& {Huchra}}{{Ebeling}
  et~al.}{1998}]{Ebeling_1998}
{Ebeling} H.,  {Edge} A.~C.,  {Bohringer} H.,  {Allen} S.~W.,  {Crawford}
  C.~S.,  {Fabian} A.~C.,  {Voges} W.,   {Huchra} J.~P.,  1998, \mn@doi
  [\mnras] {10.1046/j.1365-8711.1998.01949.x}, \href
  {https://ui.adsabs.harvard.edu/abs/1998MNRAS.301..881E} {301, 881}

\bibitem[\protect\citeauthoryear{{Ebeling}, {Edge}, {Allen}, {Crawford},
  {Fabian}  \& {Huchra}}{{Ebeling} et~al.}{2000}]{Ebeling_2000}
{Ebeling} H.,  {Edge} A.~C.,  {Allen} S.~W.,  {Crawford} C.~S.,  {Fabian}
  A.~C.,   {Huchra} J.~P.,  2000, \mn@doi [\mnras]
  {10.1046/j.1365-8711.2000.03549.x}, \href
  {https://ui.adsabs.harvard.edu/abs/2000MNRAS.318..333E} {318, 333}

\bibitem[\protect\citeauthoryear{{Ebeling}, {Stephenson}  \& {Edge}}{{Ebeling}
  et~al.}{2014}]{Ebeling2014}
{Ebeling} H.,  {Stephenson} L.~N.,   {Edge} A.~C.,  2014, \mn@doi [\apjl]
  {10.1088/2041-8205/781/2/L40}, \href
  {https://ui.adsabs.harvard.edu/abs/2014ApJ...781L..40E} {781, L40}

\bibitem[\protect\citeauthoryear{{Fabricant}, {Beers}, {Geller}, {Gorenstein},
  {Huchra}  \& {Kurtz}}{{Fabricant} et~al.}{1986}]{Fabricant1986}
{Fabricant} D.,  {Beers} T.~C.,  {Geller} M.~J.,  {Gorenstein} P.,  {Huchra}
  J.~P.,   {Kurtz} M.~J.,  1986, \mn@doi [\apj] {10.1086/164523}, \href
  {https://ui.adsabs.harvard.edu/abs/1986ApJ...308..530F} {308, 530}

\bibitem[\protect\citeauthoryear{{Fasano} et~al.,}{{Fasano}
  et~al.}{2006}]{Fasano_2006}
{Fasano} G.,  et~al., 2006, \mn@doi [\aap] {10.1051/0004-6361:20053816}, \href
  {https://ui.adsabs.harvard.edu/abs/2006A&A...445..805F} {445, 805}

\bibitem[\protect\citeauthoryear{{Fasano} et~al.,}{{Fasano}
  et~al.}{2010}]{Fasano_2010}
{Fasano} G.,  et~al., 2010, \mn@doi [\mnras]
  {10.1111/j.1365-2966.2010.16361.x}, \href
  {https://ui.adsabs.harvard.edu/abs/2010MNRAS.404.1490F} {404, 1490}

\bibitem[\protect\citeauthoryear{{Fasano} et~al.,}{{Fasano}
  et~al.}{2012}]{Fasano_2012}
{Fasano} G.,  et~al., 2012, \mn@doi [\mnras]
  {10.1111/j.1365-2966.2011.19798.x}, \href
  {https://ui.adsabs.harvard.edu/abs/2012MNRAS.420..926F} {420, 926}

\bibitem[\protect\citeauthoryear{{Fasano} et~al.,}{{Fasano}
  et~al.}{2015}]{Fasano_2015}
{Fasano} G.,  et~al., 2015, \mn@doi [\mnras] {10.1093/mnras/stv500}, \href
  {https://ui.adsabs.harvard.edu/abs/2015MNRAS.449.3927F} {449, 3927}

\bibitem[\protect\citeauthoryear{{Fujita} \& {Goto}}{{Fujita} \&
  {Goto}}{2004}]{Fujita2004}
{Fujita} Y.,  {Goto} T.,  2004, \mn@doi [\pasj] {10.1093/pasj/56.4.621}, \href
  {https://ui.adsabs.harvard.edu/abs/2004PASJ...56..621F} {56, 621}

\bibitem[\protect\citeauthoryear{{Fujita} \& {Nagashima}}{{Fujita} \&
  {Nagashima}}{1999}]{Fujita_1999}
{Fujita} Y.,  {Nagashima} M.,  1999, \mn@doi [\apj] {10.1086/307139}, \href
  {https://ui.adsabs.harvard.edu/abs/1999ApJ...516..619F} {516, 619}

\bibitem[\protect\citeauthoryear{{Fumagalli}, {Fossati}, {Hau}, {Gavazzi},
  {Bower}, {Sun}  \& {Boselli}}{{Fumagalli} et~al.}{2014}]{Fumagalli2014}
{Fumagalli} M.,  {Fossati} M.,  {Hau} G. K.~T.,  {Gavazzi} G.,  {Bower} R.,
  {Sun} M.,   {Boselli} A.,  2014, \mn@doi [\mnras] {10.1093/mnras/stu2092},
  \href {https://ui.adsabs.harvard.edu/abs/2014MNRAS.445.4335F} {445, 4335}

\bibitem[\protect\citeauthoryear{{Gebhardt}, {Pryor}, {Williams}  \&
  {Hesser}}{{Gebhardt} et~al.}{1994}]{Gebhardt1994}
{Gebhardt} K.,  {Pryor} C.,  {Williams} T.~B.,   {Hesser} J.~E.,  1994, \mn@doi
  [\aj] {10.1086/117017}, \href
  {https://ui.adsabs.harvard.edu/abs/1994AJ....107.2067G} {107, 2067}

\bibitem[\protect\citeauthoryear{{Ghizzardi}, {Rossetti}  \&
  {Molendi}}{{Ghizzardi} et~al.}{2010}]{Ghizzardi2010}
{Ghizzardi} S.,  {Rossetti} M.,   {Molendi} S.,  2010, \mn@doi [\aap]
  {10.1051/0004-6361/200912496}, \href
  {https://ui.adsabs.harvard.edu/abs/2010A&A...516A..32G} {516, A32}

\bibitem[\protect\citeauthoryear{{Ghizzardi}, {De Grandi}  \&
  {Molendi}}{{Ghizzardi} et~al.}{2013}]{Ghizzardi2013}
{Ghizzardi} S.,  {De Grandi} S.,   {Molendi} S.,  2013, \mn@doi [Astronomische
  Nachrichten] {10.1002/asna.201211871}, \href
  {https://ui.adsabs.harvard.edu/abs/2013AN....334..422G} {334, 422}

\bibitem[\protect\citeauthoryear{{Gozaliasl} et~al.,}{{Gozaliasl}
  et~al.}{2014}]{Gozaliasl2014}
{Gozaliasl} G.,  et~al., 2014, \mn@doi [\aap] {10.1051/0004-6361/201322459},
  \href {https://ui.adsabs.harvard.edu/abs/2014A&A...566A.140G} {566, A140}

\bibitem[\protect\citeauthoryear{{Gu} et~al.,}{{Gu} et~al.}{2019}]{Gu2019}
{Gu} L.,  et~al., 2019, \mn@doi [Nature Astronomy] {10.1038/s41550-019-0798-8},
  \href {https://ui.adsabs.harvard.edu/abs/2019NatAs...3..838G} {3, 838}

\bibitem[\protect\citeauthoryear{{Gullieuszik} et~al.,}{{Gullieuszik}
  et~al.}{2015}]{Gullieuszik_2015}
{Gullieuszik} M.,  et~al., 2015, \mn@doi [\aap] {10.1051/0004-6361/201526061},
  \href {https://ui.adsabs.harvard.edu/abs/2015A&A...581A..41G} {581, A41}

\bibitem[\protect\citeauthoryear{{Gullieuszik} et~al.,}{{Gullieuszik}
  et~al.}{2020}]{Gullieuszik2020}
{Gullieuszik} M.,  et~al., 2020, \mn@doi [\apj] {10.3847/1538-4357/aba3cb},
  \href {https://ui.adsabs.harvard.edu/abs/2020ApJ...899...13G} {899, 13}

\bibitem[\protect\citeauthoryear{{Gunn} \& {Gott}}{{Gunn} \&
  {Gott}}{1972}]{Gunn1972}
{Gunn} J.~E.,  {Gott} J.~Richard I.,  1972, \mn@doi [\apj] {10.1086/151605},
  \href {https://ui.adsabs.harvard.edu/abs/1972ApJ...176....1G} {176, 1}

\bibitem[\protect\citeauthoryear{{Ha}, {Ryu}  \& {Kang}}{{Ha}
  et~al.}{2018}]{Ha2018}
{Ha} J.-H.,  {Ryu} D.,   {Kang} H.,  2018, \mn@doi [\apj]
  {10.3847/1538-4357/aab4a2}, \href
  {https://ui.adsabs.harvard.edu/abs/2018ApJ...857...26H} {857, 26}

\bibitem[\protect\citeauthoryear{{Haines} et~al.,}{{Haines}
  et~al.}{2023}]{Haines2023}
{Haines} C.,  et~al., 2023, \mn@doi [The Messenger] {10.18727/0722-6691/5308},
  \href {https://ui.adsabs.harvard.edu/abs/2023Msngr.190...31H} {190, 31}

\bibitem[\protect\citeauthoryear{{Hallman} \& {Markevitch}}{{Hallman} \&
  {Markevitch}}{2004}]{Hallman2004}
{Hallman} E.~J.,  {Markevitch} M.,  2004, \mn@doi [\apjl] {10.1086/423449},
  \href {https://ui.adsabs.harvard.edu/abs/2004ApJ...610L..81H} {610, L81}

\bibitem[\protect\citeauthoryear{{Hallman}, {Skillman}, {Jeltema}, {Smith},
  {O'Shea}, {Burns}  \& {Norman}}{{Hallman} et~al.}{2010}]{Hallman2010}
{Hallman} E.~J.,  {Skillman} S.~W.,  {Jeltema} T.~E.,  {Smith} B.~D.,  {O'Shea}
  B.~W.,  {Burns} J.~O.,   {Norman} M.~L.,  2010, \mn@doi [\apj]
  {10.1088/0004-637X/725/1/1053}, \href
  {https://ui.adsabs.harvard.edu/abs/2010ApJ...725.1053H} {725, 1053}

\bibitem[\protect\citeauthoryear{{Haynes} et~al.,}{{Haynes}
  et~al.}{2018}]{Haynes2018}
{Haynes} M.~P.,  et~al., 2018, \mn@doi [\apj] {10.3847/1538-4357/aac956}, \href
  {https://ui.adsabs.harvard.edu/abs/2018ApJ...861...49H} {861, 49}

\bibitem[\protect\citeauthoryear{{Hudson}, {Mittal}, {Reiprich}, {Nulsen},
  {Andernach}  \& {Sarazin}}{{Hudson} et~al.}{2010}]{Hudson_2010}
{Hudson} D.~S.,  {Mittal} R.,  {Reiprich} T.~H.,  {Nulsen} P.~E.~J.,
  {Andernach} H.,   {Sarazin} C.~L.,  2010, \mn@doi [\aap]
  {10.1051/0004-6361/200912377}, \href
  {https://ui.adsabs.harvard.edu/abs/2010A&A...513A..37H} {513, A37}

\bibitem[\protect\citeauthoryear{{Hwang} \& {Lee}}{{Hwang} \&
  {Lee}}{2009}]{Hwang2009}
{Hwang} H.~S.,  {Lee} M.~G.,  2009, \mn@doi [\mnras]
  {10.1111/j.1365-2966.2009.15100.x}, \href
  {https://ui.adsabs.harvard.edu/abs/2009MNRAS.397.2111H} {397, 2111}

\bibitem[\protect\citeauthoryear{{Ichinohe}, {Werner}, {Simionescu}, {Allen},
  {Canning}, {Ehlert}, {Mernier}  \& {Takahashi}}{{Ichinohe}
  et~al.}{2015}]{Ichinohe2015}
{Ichinohe} Y.,  {Werner} N.,  {Simionescu} A.,  {Allen} S.~W.,  {Canning}
  R.~E.~A.,  {Ehlert} S.,  {Mernier} F.,   {Takahashi} T.,  2015, \mn@doi
  [\mnras] {10.1093/mnras/stv217}, \href
  {https://ui.adsabs.harvard.edu/abs/2015MNRAS.448.2971I} {448, 2971}

\bibitem[\protect\citeauthoryear{{Ignesti}, {Gitti}, {Brunetti}, {O'Sullivan},
  {Sarazin}  \& {Wong}}{{Ignesti} et~al.}{2018}]{Ignesti2018}
{Ignesti} A.,  {Gitti} M.,  {Brunetti} G.,  {O'Sullivan} E.,  {Sarazin} C.,
  {Wong} K.,  2018, \mn@doi [\aap] {10.1051/0004-6361/201731380}, \href
  {https://ui.adsabs.harvard.edu/abs/2018A&A...610A..89I} {610, A89}

\bibitem[\protect\citeauthoryear{{Ignesti} et~al.,}{{Ignesti}
  et~al.}{2022}]{Ignesti_2022}
{Ignesti} A.,  et~al., 2022, \mn@doi [\apj] {10.3847/1538-4357/ac8cf6}, \href
  {https://ui.adsabs.harvard.edu/abs/2022ApJ...937...58I} {937, 58}

\bibitem[\protect\citeauthoryear{{Ignesti} et~al.,}{{Ignesti}
  et~al.}{2023}]{Ignesti2023}
{Ignesti} A.,  et~al., 2023, \mn@doi [arXiv e-prints]
  {10.48550/arXiv.2305.19941}, \href
  {https://ui.adsabs.harvard.edu/abs/2023arXiv230519941I} {p. arXiv:2305.19941}

\bibitem[\protect\citeauthoryear{{Inoue}, {Hayashida}, {Ueda}, {Nagino},
  {Tsunemi}  \& {Koyama}}{{Inoue} et~al.}{2016}]{Inoue2016}
{Inoue} S.,  {Hayashida} K.,  {Ueda} S.,  {Nagino} R.,  {Tsunemi} H.,
  {Koyama} K.,  2016, \mn@doi [\pasj] {10.1093/pasj/psw027}, \href
  {https://ui.adsabs.harvard.edu/abs/2016PASJ...68S..23I} {68, S23}

\bibitem[\protect\citeauthoryear{{Jaff{\'e}}, {Poggianti}, {Verheijen},
  {Deshev}  \& {van Gorkom}}{{Jaff{\'e}} et~al.}{2013}]{Jaffe_2013}
{Jaff{\'e}} Y.~L.,  {Poggianti} B.~M.,  {Verheijen} M. A.~W.,  {Deshev} B.~Z.,
   {van Gorkom} J.~H.,  2013, \mn@doi [\mnras] {10.1093/mnras/stt250}, \href
  {https://ui.adsabs.harvard.edu/abs/2013MNRAS.431.2111J} {431, 2111}

\bibitem[\protect\citeauthoryear{{Jaff{\'e}}, {Smith}, {Candlish}, {Poggianti},
  {Sheen}  \& {Verheijen}}{{Jaff{\'e}} et~al.}{2015}]{Jaffe_2015}
{Jaff{\'e}} Y.~L.,  {Smith} R.,  {Candlish} G.~N.,  {Poggianti} B.~M.,  {Sheen}
  Y.-K.,   {Verheijen} M. A.~W.,  2015, \mn@doi [\mnras]
  {10.1093/mnras/stv100}, \href
  {https://ui.adsabs.harvard.edu/abs/2015MNRAS.448.1715J} {448, 1715}

\bibitem[\protect\citeauthoryear{{Jaff{\'e}} et~al.,}{{Jaff{\'e}}
  et~al.}{2018}]{Jaffe2018}
{Jaff{\'e}} Y.~L.,  et~al., 2018, \mn@doi [\mnras] {10.1093/mnras/sty500},
  \href {https://ui.adsabs.harvard.edu/abs/2018MNRAS.476.4753J} {476, 4753}

\bibitem[\protect\citeauthoryear{{Jin} et~al.,}{{Jin} et~al.}{2023}]{Jin2023}
{Jin} S.,  et~al., 2023, \mn@doi [\mnras] {10.1093/mnras/stad557}, \href
  {https://ui.adsabs.harvard.edu/abs/2023MNRAS.tmp..715J} {}

\bibitem[\protect\citeauthoryear{{Jones}, {Ponman}, {Horton}, {Babul},
  {Ebeling}  \& {Burke}}{{Jones} et~al.}{2003}]{Jones_2003}
{Jones} L.~R.,  {Ponman} T.~J.,  {Horton} A.,  {Babul} A.,  {Ebeling} H.,
  {Burke} D.~J.,  2003, \mn@doi [\mnras] {10.1046/j.1365-8711.2003.06702.x},
  \href {https://ui.adsabs.harvard.edu/abs/2003MNRAS.343..627J} {343, 627}

\bibitem[\protect\citeauthoryear{{Kapferer}, {Sluka}, {Schindler}, {Ferrari}
  \& {Ziegler}}{{Kapferer} et~al.}{2009}]{Kapferer_2009}
{Kapferer} W.,  {Sluka} C.,  {Schindler} S.,  {Ferrari} C.,   {Ziegler} B.,
  2009, \mn@doi [\aap] {10.1051/0004-6361/200811551}, \href
  {https://ui.adsabs.harvard.edu/abs/2009A&A...499...87K} {499, 87}

\bibitem[\protect\citeauthoryear{{Kass} \& {Wasserman}}{{Kass} \&
  {Wasserman}}{1995}]{Kass1995}
{Kass} R.~E.,  {Wasserman} L.,  1995, Journal of the American Statistical
  Association, 90, 928

\bibitem[\protect\citeauthoryear{{Katayama}, {Hayashida}, {Takahara}  \&
  {Fujita}}{{Katayama} et~al.}{2003}]{Katayama_2003}
{Katayama} H.,  {Hayashida} K.,  {Takahara} F.,   {Fujita} Y.,  2003, \mn@doi
  [\apj] {10.1086/346126}, \href
  {https://ui.adsabs.harvard.edu/abs/2003ApJ...585..687K} {585, 687}

\bibitem[\protect\citeauthoryear{{Kelkar} et~al.,}{{Kelkar}
  et~al.}{2020}]{Kelkar_2020}
{Kelkar} K.,  et~al., 2020, arXiv e-prints, \href
  {https://ui.adsabs.harvard.edu/abs/2020arXiv200208610K} {p. arXiv:2002.08610}

\bibitem[\protect\citeauthoryear{{Knopp}, {Henry}  \& {Briel}}{{Knopp}
  et~al.}{1996}]{Knopp1996}
{Knopp} G.~P.,  {Henry} J.~P.,   {Briel} U.~G.,  1996, \mn@doi [\apj]
  {10.1086/178047}, \href
  {https://ui.adsabs.harvard.edu/abs/1996ApJ...472..125K} {472, 125}

\bibitem[\protect\citeauthoryear{{Kolcu}, {Crossett}, {Bellhouse}  \&
  {McGee}}{{Kolcu} et~al.}{2022}]{Kolcu2022}
{Kolcu} T.,  {Crossett} J.~P.,  {Bellhouse} C.,   {McGee} S.,  2022, \mn@doi
  [\mnras] {10.1093/mnras/stac2177}, \href
  {https://ui.adsabs.harvard.edu/abs/2022MNRAS.515.5877K} {515, 5877}

\bibitem[\protect\citeauthoryear{Kruskal \& Wallis}{Kruskal \&
  Wallis}{1952}]{Kruskal1952}
Kruskal W.~H.,  Wallis W.~A.,  1952, Journal of the American Statistical
  Association, 47, 583

\bibitem[\protect\citeauthoryear{{Lagan{\'a}}, {Andrade-Santos}  \& {Lima
  Neto}}{{Lagan{\'a}} et~al.}{2010}]{Lagana2010}
{Lagan{\'a}} T.~F.,  {Andrade-Santos} F.,   {Lima Neto} G.~B.,  2010, \mn@doi
  [\aap] {10.1051/0004-6361/200913180}, \href
  {https://ui.adsabs.harvard.edu/abs/2010A&A...511A..15L} {511, A15}

\bibitem[\protect\citeauthoryear{{Lagan{\'a}}, {Durret}  \&
  {Lopes}}{{Lagan{\'a}} et~al.}{2019}]{Lagana2019}
{Lagan{\'a}} T.~F.,  {Durret} F.,   {Lopes} P.~A.~A.,  2019, \mn@doi [\mnras]
  {10.1093/mnras/stz148}, \href
  {https://ui.adsabs.harvard.edu/abs/2019MNRAS.484.2807L} {484, 2807}

\bibitem[\protect\citeauthoryear{{Lakhchaura}, {Singh}, {Saikia}  \&
  {Hunstead}}{{Lakhchaura} et~al.}{2013}]{Lakhchaura2013}
{Lakhchaura} K.,  {Singh} K.~P.,  {Saikia} D.~J.,   {Hunstead} R.~W.,  2013,
  \mn@doi [\apj] {10.1088/0004-637X/767/1/91}, \href
  {https://ui.adsabs.harvard.edu/abs/2013ApJ...767...91L} {767, 91}

\bibitem[\protect\citeauthoryear{{Larson}, {Tinsley}  \& {Caldwell}}{{Larson}
  et~al.}{1980}]{Larson1980}
{Larson} R.~B.,  {Tinsley} B.~M.,   {Caldwell} C.~N.,  1980, \mn@doi [\apj]
  {10.1086/157917}, \href
  {https://ui.adsabs.harvard.edu/abs/1980ApJ...237..692L} {237, 692}

\bibitem[\protect\citeauthoryear{{Laudari} et~al.,}{{Laudari}
  et~al.}{2022}]{Laudari_2022}
{Laudari} S.,  et~al., 2022, \mn@doi [\mnras] {10.1093/mnras/stab3280}, \href
  {https://ui.adsabs.harvard.edu/abs/2022MNRAS.509.3938L} {509, 3938}

\bibitem[\protect\citeauthoryear{{Liske}, {Lemon}, {Driver}, {Cross}  \&
  {Couch}}{{Liske} et~al.}{2003}]{Liske2003}
{Liske} J.,  {Lemon} D.~J.,  {Driver} S.~P.,  {Cross} N.~J.~G.,   {Couch}
  W.~J.,  2003, \mn@doi [\mnras] {10.1046/j.1365-8711.2003.06826.x}, \href
  {https://ui.adsabs.harvard.edu/abs/2003MNRAS.344..307L} {344, 307}

\bibitem[\protect\citeauthoryear{{Lopes}, {Trevisan}, {Lagan{\'a}}, {Durret},
  {Ribeiro}  \& {Rembold}}{{Lopes} et~al.}{2018}]{Lopes_2018}
{Lopes} P. A.~A.,  {Trevisan} M.,  {Lagan{\'a}} T.~F.,  {Durret} F.,  {Ribeiro}
  A.~L.~B.,   {Rembold} S.~B.,  2018, \mn@doi [\mnras] {10.1093/mnras/sty1374},
  \href {https://ui.adsabs.harvard.edu/abs/2018MNRAS.478.5473L} {478, 5473}

\bibitem[\protect\citeauthoryear{{Louren{\c{c}}o} et~al.,}{{Louren{\c{c}}o}
  et~al.}{2020}]{Lourenco2020}
{Louren{\c{c}}o} A. C.~C.,  et~al., 2020, \mn@doi [\mnras]
  {10.1093/mnras/staa2464}, \href
  {https://ui.adsabs.harvard.edu/abs/2020MNRAS.498..835L} {498, 835}

\bibitem[\protect\citeauthoryear{{Ma}, {Ebeling}, {Marshall}  \&
  {Schrabback}}{{Ma} et~al.}{2010}]{Ma2010}
{Ma} C.~J.,  {Ebeling} H.,  {Marshall} P.,   {Schrabback} T.,  2010, \mn@doi
  [\mnras] {10.1111/j.1365-2966.2010.16673.x}, \href
  {https://ui.adsabs.harvard.edu/abs/2010MNRAS.406..121M} {406, 121}

\bibitem[\protect\citeauthoryear{{Macario}, {Markevitch}, {Giacintucci},
  {Brunetti}, {Venturi}  \& {Murray}}{{Macario} et~al.}{2011}]{Macario_2011}
{Macario} G.,  {Markevitch} M.,  {Giacintucci} S.,  {Brunetti} G.,  {Venturi}
  T.,   {Murray} S.~S.,  2011, \mn@doi [\apj] {10.1088/0004-637X/728/2/82},
  \href {https://ui.adsabs.harvard.edu/abs/2011ApJ...728...82M} {728, 82}

\bibitem[\protect\citeauthoryear{{Machado} \& {Lima Neto}}{{Machado} \& {Lima
  Neto}}{2013}]{Machado2013}
{Machado} R.~E.~G.,  {Lima Neto} G.~B.,  2013, \mn@doi [\mnras]
  {10.1093/mnras/stt127}, \href
  {http://adsabs.harvard.edu/abs/2013MNRAS.430.3249M} {430, 3249}

\bibitem[\protect\citeauthoryear{{Mann} \& {Ebeling}}{{Mann} \&
  {Ebeling}}{2012}]{Mann_ebeling2012}
{Mann} A.~W.,  {Ebeling} H.,  2012, \mn@doi [\mnras]
  {10.1111/j.1365-2966.2011.20170.x}, \href
  {https://ui.adsabs.harvard.edu/abs/2012MNRAS.420.2120M} {420, 2120}

\bibitem[\protect\citeauthoryear{{Markevitch} \& {Vikhlinin}}{{Markevitch} \&
  {Vikhlinin}}{2007}]{Markevitch2007}
{Markevitch} M.,  {Vikhlinin} A.,  2007, \mn@doi [\physrep]
  {10.1016/j.physrep.2007.01.001}, \href
  {https://ui.adsabs.harvard.edu/abs/2007PhR...443....1M} {443, 1}

\bibitem[\protect\citeauthoryear{{Markevitch} et~al.,}{{Markevitch}
  et~al.}{2000}]{Markevitch_2000}
{Markevitch} M.,  et~al., 2000, \mn@doi [\apj] {10.1086/309470}, \href
  {https://ui.adsabs.harvard.edu/abs/2000ApJ...541..542M} {541, 542}

\bibitem[\protect\citeauthoryear{{Mazzotta}, {Fusco-Femiano}  \&
  {Vikhlinin}}{{Mazzotta} et~al.}{2002}]{Mazzotta2002}
{Mazzotta} P.,  {Fusco-Femiano} R.,   {Vikhlinin} A.,  2002, \mn@doi [\apjl]
  {10.1086/340481}, \href
  {https://ui.adsabs.harvard.edu/abs/2002ApJ...569L..31M} {569, L31}

\bibitem[\protect\citeauthoryear{{McGee}, {Balogh}, {Bower}, {Font}  \&
  {McCarthy}}{{McGee} et~al.}{2009}]{McGee_2009}
{McGee} S.~L.,  {Balogh} M.~L.,  {Bower} R.~G.,  {Font} A.~S.,   {McCarthy}
  I.~G.,  2009, \mn@doi [\mnras] {10.1111/j.1365-2966.2009.15507.x}, \href
  {https://ui.adsabs.harvard.edu/abs/2009MNRAS.400..937M} {400, 937}

\bibitem[\protect\citeauthoryear{{McPartland}, {Ebeling}, {Roediger}  \&
  {Blumenthal}}{{McPartland} et~al.}{2016}]{McPartland2016}
{McPartland} C.,  {Ebeling} H.,  {Roediger} E.,   {Blumenthal} K.,  2016,
  \mn@doi [\mnras] {10.1093/mnras/stv2508}, \href
  {https://ui.adsabs.harvard.edu/abs/2016MNRAS.455.2994M} {455, 2994}

\bibitem[\protect\citeauthoryear{{Merloni} et~al.,}{{Merloni}
  et~al.}{2012}]{Merloni2012}
{Merloni} A.,  et~al., 2012, \mn@doi [arXiv e-prints]
  {10.48550/arXiv.1209.3114}, \href
  {https://ui.adsabs.harvard.edu/abs/2012arXiv1209.3114M} {p. arXiv:1209.3114}

\bibitem[\protect\citeauthoryear{{Mitsuishi}, {Babazaki}, {Ota}, {Sasaki},
  {B{\"o}hringer}, {Chon}  \& {Pratt}}{{Mitsuishi}
  et~al.}{2018}]{Mitsuishi2018}
{Mitsuishi} I.,  {Babazaki} Y.,  {Ota} N.,  {Sasaki} S.,  {B{\"o}hringer} H.,
  {Chon} G.,   {Pratt} G.~W.,  2018, \mn@doi [\pasj] {10.1093/pasj/psy117},
  \href {https://ui.adsabs.harvard.edu/abs/2018PASJ...70..112M} {70, 112}

\bibitem[\protect\citeauthoryear{{Monteiro-Oliveira}, {Lima Neto}, {Cypriano},
  {Machado}, {Capelato}, {Lagan{\'a}}, {Durret}  \&
  {Bagchi}}{{Monteiro-Oliveira} et~al.}{2017}]{Monteiro2017b}
{Monteiro-Oliveira} R.,  {Lima Neto} G.~B.,  {Cypriano} E.~S.,  {Machado}
  R.~E.~G.,  {Capelato} H.~V.,  {Lagan{\'a}} T.~F.,  {Durret} F.,   {Bagchi}
  J.,  2017, \mn@doi [\mnras] {10.1093/mnras/stx791}, \href
  {https://ui.adsabs.harvard.edu/abs/2017MNRAS.468.4566M} {468, 4566}

\bibitem[\protect\citeauthoryear{{Moore}, {Katz}, {Lake}, {Dressler}  \&
  {Oemler}}{{Moore} et~al.}{1996}]{Moore1996}
{Moore} B.,  {Katz} N.,  {Lake} G.,  {Dressler} A.,   {Oemler} A.,  1996,
  \mn@doi [\nat] {10.1038/379613a0}, \href
  {https://ui.adsabs.harvard.edu/abs/1996Natur.379..613M} {379, 613}

\bibitem[\protect\citeauthoryear{{Moretti} et~al.,}{{Moretti}
  et~al.}{2014}]{Moretti2014}
{Moretti} A.,  et~al., 2014, \mn@doi [\aap] {10.1051/0004-6361/201323098},
  \href {https://ui.adsabs.harvard.edu/abs/2014A&A...564A.138M} {564, A138}

\bibitem[\protect\citeauthoryear{{Moretti} et~al.,}{{Moretti}
  et~al.}{2017}]{Moretti2017}
{Moretti} A.,  et~al., 2017, \mn@doi [\aap] {10.1051/0004-6361/201630030},
  \href {https://ui.adsabs.harvard.edu/abs/2017A&A...599A..81M} {599, A81}

\bibitem[\protect\citeauthoryear{{Munari}, {Biviano}, {Borgani}, {Murante}  \&
  {Fabjan}}{{Munari} et~al.}{2013}]{Munari_2013}
{Munari} E.,  {Biviano} A.,  {Borgani} S.,  {Murante} G.,   {Fabjan} D.,  2013,
  \mn@doi [\mnras] {10.1093/mnras/stt049}, \href
  {https://ui.adsabs.harvard.edu/abs/2013MNRAS.430.2638M} {430, 2638}

\bibitem[\protect\citeauthoryear{{Nakazawa} et~al.,}{{Nakazawa}
  et~al.}{2009}]{Nakazawa2009}
{Nakazawa} K.,  et~al., 2009, \mn@doi [\pasj] {10.1093/pasj/61.2.339}, \href
  {https://ui.adsabs.harvard.edu/abs/2009PASJ...61..339N} {61, 339}

\bibitem[\protect\citeauthoryear{{Navarro}, {Frenk}  \& {White}}{{Navarro}
  et~al.}{1997}]{Navarro1997}
{Navarro} J.~F.,  {Frenk} C.~S.,   {White} S. D.~M.,  1997, \mn@doi [\apj]
  {10.1086/304888}, \href
  {https://ui.adsabs.harvard.edu/abs/1997ApJ...490..493N} {490, 493}

\bibitem[\protect\citeauthoryear{{Nulsen}}{{Nulsen}}{1982}]{Nulsen1982}
{Nulsen} P.~E.~J.,  1982, \mn@doi [\mnras] {10.1093/mnras/198.4.1007}, \href
  {https://ui.adsabs.harvard.edu/abs/1982MNRAS.198.1007N} {198, 1007}

\bibitem[\protect\citeauthoryear{{Oman}, {Bah{\'e}}, {Healy}, {Hess}, {Hudson}
  \& {Verheijen}}{{Oman} et~al.}{2021}]{Oman_2021}
{Oman} K.~A.,  {Bah{\'e}} Y.~M.,  {Healy} J.,  {Hess} K.~M.,  {Hudson} M.~J.,
  {Verheijen} M. A.~W.,  2021, \mn@doi [\mnras] {10.1093/mnras/staa3845}, \href
  {https://ui.adsabs.harvard.edu/abs/2021MNRAS.501.5073O} {501, 5073}

\bibitem[\protect\citeauthoryear{{Owers}, {Nulsen}, {Couch}  \&
  {Markevitch}}{{Owers} et~al.}{2009}]{Owers_2009}
{Owers} M.~S.,  {Nulsen} P. E.~J.,  {Couch} W.~J.,   {Markevitch} M.,  2009,
  \mn@doi [\apj] {10.1088/0004-637X/704/2/1349}, \href
  {https://ui.adsabs.harvard.edu/abs/2009ApJ...704.1349O} {704, 1349}

\bibitem[\protect\citeauthoryear{{Owers}, {Couch}, {Nulsen}  \& {Rand
  all}}{{Owers} et~al.}{2012}]{Owers2012}
{Owers} M.~S.,  {Couch} W.~J.,  {Nulsen} P. E.~J.,   {Rand all} S.~W.,  2012,
  \mn@doi [\apjl] {10.1088/2041-8205/750/1/L23}, \href
  {https://ui.adsabs.harvard.edu/abs/2012ApJ...750L..23O} {750, L23}

\bibitem[\protect\citeauthoryear{{Paccagnella} et~al.,}{{Paccagnella}
  et~al.}{2017}]{Paccagnella_2017}
{Paccagnella} A.,  et~al., 2017, \mn@doi [\apj] {10.3847/1538-4357/aa64d7},
  \href {https://ui.adsabs.harvard.edu/abs/2017ApJ...838..148P} {838, 148}

\bibitem[\protect\citeauthoryear{{Pallero}, {G{\'o}mez}, {Padilla},
  {Torres-Flores}, {Demarco}, {Cerulo}  \& {Olave-Rojas}}{{Pallero}
  et~al.}{2019}]{Pallero2019}
{Pallero} D.,  {G{\'o}mez} F.~A.,  {Padilla} N.~D.,  {Torres-Flores} S.,
  {Demarco} R.,  {Cerulo} P.,   {Olave-Rojas} D.,  2019, \mn@doi [\mnras]
  {10.1093/mnras/stz1745}, \href
  {https://ui.adsabs.harvard.edu/abs/2019MNRAS.488..847P} {488, 847}

\bibitem[\protect\citeauthoryear{{Pallero}, {G{\'o}mez}, {Padilla}, {Bah{\'e}},
  {Vega-Mart{\'\i}nez}  \& {Torres-Flores}}{{Pallero}
  et~al.}{2022}]{Pallero2022}
{Pallero} D.,  {G{\'o}mez} F.~A.,  {Padilla} N.~D.,  {Bah{\'e}} Y.~M.,
  {Vega-Mart{\'\i}nez} C.~A.,   {Torres-Flores} S.,  2022, \mn@doi [\mnras]
  {10.1093/mnras/stab3318}, \href
  {https://ui.adsabs.harvard.edu/abs/2022MNRAS.511.3210P} {511, 3210}

\bibitem[\protect\citeauthoryear{{Peebles}}{{Peebles}}{1993}]{Peebles1993}
{Peebles} P.~J.~E.,  1993, {Principles of Physical Cosmology}

\bibitem[\protect\citeauthoryear{{Poggianti}}{{Poggianti}}{1997}]{Poggianti1997}
{Poggianti} B.~M.,  1997, \mn@doi [\aaps] {10.1051/aas:1997142}, \href
  {https://ui.adsabs.harvard.edu/abs/1997A&AS..122..399P} {122, 399}

\bibitem[\protect\citeauthoryear{{Poggianti} et~al.,}{{Poggianti}
  et~al.}{2016}]{Poggianti2016}
{Poggianti} B.~M.,  et~al., 2016, \mn@doi [\aj] {10.3847/0004-6256/151/3/78},
  \href {https://ui.adsabs.harvard.edu/abs/2016AJ....151...78P} {151, 78}

\bibitem[\protect\citeauthoryear{{Poggianti} et~al.,}{{Poggianti}
  et~al.}{2017}]{Poggianti_2017}
{Poggianti} B.~M.,  et~al., 2017, \mn@doi [\apj] {10.3847/1538-4357/aa78ed},
  \href {https://ui.adsabs.harvard.edu/abs/2017ApJ...844...48P} {844, 48}

\bibitem[\protect\citeauthoryear{{Poggianti} et~al.,}{{Poggianti}
  et~al.}{2019}]{Poggianti_2019}
{Poggianti} B.~M.,  et~al., 2019, \mn@doi [\mnras] {10.1093/mnras/sty2999},
  \href {https://ui.adsabs.harvard.edu/abs/2019MNRAS.482.4466P} {482, 4466}

\bibitem[\protect\citeauthoryear{{Poole}, {Fardal}, {Babul}, {McCarthy},
  {Quinn}  \& {Wadsley}}{{Poole} et~al.}{2006}]{Poole2006}
{Poole} G.~B.,  {Fardal} M.~A.,  {Babul} A.,  {McCarthy} I.~G.,  {Quinn} T.,
  {Wadsley} J.,  2006, \mn@doi [\mnras] {10.1111/j.1365-2966.2006.10916.x},
  \href {https://ui.adsabs.harvard.edu/abs/2006MNRAS.373..881P} {373, 881}

\bibitem[\protect\citeauthoryear{{Press} \& {Schechter}}{{Press} \&
  {Schechter}}{1974}]{PressSchechter1974}
{Press} W.~H.,  {Schechter} P.,  1974, \mn@doi [\apj] {10.1086/152650}, \href
  {https://ui.adsabs.harvard.edu/abs/1974ApJ...187..425P} {187, 425}

\bibitem[\protect\citeauthoryear{{Randall}, {Clarke}, {Nulsen}, {Owers},
  {Sarazin}, {Forman}  \& {Murray}}{{Randall} et~al.}{2010}]{Randall2010}
{Randall} S.~W.,  {Clarke} T.~E.,  {Nulsen} P.~E.~J.,  {Owers} M.~S.,
  {Sarazin} C.~L.,  {Forman} W.~R.,   {Murray} S.~S.,  2010, \mn@doi [\apj]
  {10.1088/0004-637X/722/1/825}, \href
  {https://ui.adsabs.harvard.edu/abs/2010ApJ...722..825R} {722, 825}

\bibitem[\protect\citeauthoryear{{Raouf}, {Smith}, {Khosroshahi}, {Dariush},
  {Driver}, {Ko}  \& {Hwang}}{{Raouf} et~al.}{2019}]{Raouf2019}
{Raouf} M.,  {Smith} R.,  {Khosroshahi} H.~G.,  {Dariush} A.~A.,  {Driver} S.,
  {Ko} J.,   {Hwang} H.~S.,  2019, \mn@doi [\apj] {10.3847/1538-4357/ab5581},
  \href {https://ui.adsabs.harvard.edu/abs/2019ApJ...887..264R} {887, 264}

\bibitem[\protect\citeauthoryear{{Rasia}, {Borgani}, {Ettori}, {Mazzotta}  \&
  {Meneghetti}}{{Rasia} et~al.}{2013}]{Rasia2013}
{Rasia} E.,  {Borgani} S.,  {Ettori} S.,  {Mazzotta} P.,   {Meneghetti} M.,
  2013, \mn@doi [\apj] {10.1088/0004-637X/776/1/39}, \href
  {https://ui.adsabs.harvard.edu/abs/2013ApJ...776...39R} {776, 39}

\bibitem[\protect\citeauthoryear{{Rawle} et~al.,}{{Rawle}
  et~al.}{2014}]{Rawle2014}
{Rawle} T.~D.,  et~al., 2014, \mn@doi [\mnras] {10.1093/mnras/stu868}, \href
  {https://ui.adsabs.harvard.edu/abs/2014MNRAS.442..196R} {442, 196}

\bibitem[\protect\citeauthoryear{{Raychaudhury}, {Fabian}, {Edge}, {Jones}  \&
  {Forman}}{{Raychaudhury} et~al.}{1991}]{Raychaudhury1991}
{Raychaudhury} S.,  {Fabian} A.~C.,  {Edge} A.~C.,  {Jones} C.,   {Forman} W.,
  1991, \mn@doi [\mnras] {10.1093/mnras/248.1.101}, \href
  {https://ui.adsabs.harvard.edu/abs/1991MNRAS.248..101R} {248, 101}

\bibitem[\protect\citeauthoryear{{Reid}, {Hunstead}  \& {Pierre}}{{Reid}
  et~al.}{1998}]{Reid1998}
{Reid} A.~D.,  {Hunstead} R.~W.,   {Pierre} M.~M.,  1998, \mn@doi [\mnras]
  {10.1046/j.1365-8711.1998.01331.x}, \href
  {https://ui.adsabs.harvard.edu/abs/1998MNRAS.296..531R} {296, 531}

\bibitem[\protect\citeauthoryear{Riseley et~al.,}{Riseley
  et~al.}{2022}]{Riseley2022}
Riseley C.~J.,  et~al., 2022, \mn@doi [Monthly Notices of the Royal
  Astronomical Society] {10.1093/mnras/stac1771}, 515, 1871

\bibitem[\protect\citeauthoryear{{Roberts}, {Parker}  \&
  {Hlavacek-Larrondo}}{{Roberts} et~al.}{2018}]{Roberts_2018}
{Roberts} I.~D.,  {Parker} L.~C.,   {Hlavacek-Larrondo} J.,  2018, \mn@doi
  [\mnras] {10.1093/mnras/sty131}, \href
  {https://ui.adsabs.harvard.edu/abs/2018MNRAS.475.4704R} {475, 4704}

\bibitem[\protect\citeauthoryear{{Roberts} et~al.,}{{Roberts}
  et~al.}{2021a}]{Roberts2021a}
{Roberts} I.~D.,  et~al., 2021a, \mn@doi [\aap] {10.1051/0004-6361/202140784},
  \href {https://ui.adsabs.harvard.edu/abs/2021A&A...650A.111R} {650, A111}

\bibitem[\protect\citeauthoryear{{Roberts}, {van Weeren}, {McGee}, {Botteon},
  {Ignesti}  \& {Rottgering}}{{Roberts} et~al.}{2021b}]{Roberts2021b}
{Roberts} I.~D.,  {van Weeren} R.~J.,  {McGee} S.~L.,  {Botteon} A.,  {Ignesti}
  A.,   {Rottgering} H.~J.~A.,  2021b, \mn@doi [\aap]
  {10.1051/0004-6361/202141118}, \href
  {https://ui.adsabs.harvard.edu/abs/2021A&A...652A.153R} {652, A153}

\bibitem[\protect\citeauthoryear{{Roediger}, {Br{\"u}ggen}, {Simionescu},
  {B{\"o}hringer}, {Churazov}  \& {Forman}}{{Roediger}
  et~al.}{2011}]{Roediger2011}
{Roediger} E.,  {Br{\"u}ggen} M.,  {Simionescu} A.,  {B{\"o}hringer} H.,
  {Churazov} E.,   {Forman} W.~R.,  2011, \mn@doi [\mnras]
  {10.1111/j.1365-2966.2011.18279.x}, \href
  {https://ui.adsabs.harvard.edu/abs/2011MNRAS.413.2057R} {413, 2057}

\bibitem[\protect\citeauthoryear{{Roettiger}, {Burns}  \& {Stone}}{{Roettiger}
  et~al.}{1999}]{Roettiger1999}
{Roettiger} K.,  {Burns} J.~O.,   {Stone} J.~M.,  1999, \mn@doi [\apj]
  {10.1086/307327}, \href
  {https://ui.adsabs.harvard.edu/abs/1999ApJ...518..603R} {518, 603}

\bibitem[\protect\citeauthoryear{{Roman-Oliveira}, {Chies-Santos},
  {Rodr{\'\i}guez del Pino}, {Arag{\'o}n-Salamanca}, {Gray}  \&
  {Bamford}}{{Roman-Oliveira} et~al.}{2019}]{RomanOliveira2019}
{Roman-Oliveira} F.~V.,  {Chies-Santos} A.~L.,  {Rodr{\'\i}guez del Pino} B.,
  {Arag{\'o}n-Salamanca} A.,  {Gray} M.~E.,   {Bamford} S.~P.,  2019, \mn@doi
  [\mnras] {10.1093/mnras/stz007}, \href
  {https://ui.adsabs.harvard.edu/abs/2019MNRAS.484..892R} {484, 892}

\bibitem[\protect\citeauthoryear{{Rossetti}, {Ghizzardi}, {Molendi}  \&
  {Finoguenov}}{{Rossetti} et~al.}{2007}]{Rossetti2007}
{Rossetti} M.,  {Ghizzardi} S.,  {Molendi} S.,   {Finoguenov} A.,  2007,
  \mn@doi [\aap] {10.1051/0004-6361:20054621}, \href
  {https://ui.adsabs.harvard.edu/abs/2007A&A...463..839R} {463, 839}

\bibitem[\protect\citeauthoryear{{Rottgering}, {Wieringa}, {Hunstead}  \&
  {Ekers}}{{Rottgering} et~al.}{1997}]{Rottgering1997}
{Rottgering} H.~J.~A.,  {Wieringa} M.~H.,  {Hunstead} R.~W.,   {Ekers} R.~D.,
  1997, \mn@doi [\mnras] {10.1093/mnras/290.4.577}, \href
  {https://ui.adsabs.harvard.edu/abs/1997MNRAS.290..577R} {290, 577}

\bibitem[\protect\citeauthoryear{{Ruggiero}, {Machado}, {Roman-Oliveira},
  {Chies-Santos}, {Lima Neto}, {Doubrawa}  \& {Rodr{\'\i}guez del
  Pino}}{{Ruggiero} et~al.}{2019}]{Ruggiero2019}
{Ruggiero} R.,  {Machado} R. E.~G.,  {Roman-Oliveira} F.~V.,  {Chies-Santos}
  A.~L.,  {Lima Neto} G.~B.,  {Doubrawa} L.,   {Rodr{\'\i}guez del Pino} B.,
  2019, \mn@doi [\mnras] {10.1093/mnras/sty3422}, \href
  {https://ui.adsabs.harvard.edu/\#abs/2019MNRAS.484..906R} {484, 906}

\bibitem[\protect\citeauthoryear{{Sanderson}, {Edge}  \& {Smith}}{{Sanderson}
  et~al.}{2009}]{Sanderson_2009}
{Sanderson} A. J.~R.,  {Edge} A.~C.,   {Smith} G.~P.,  2009, \mn@doi [\mnras]
  {10.1111/j.1365-2966.2009.15214.x}, \href
  {https://ui.adsabs.harvard.edu/abs/2009MNRAS.398.1698S} {398, 1698}

\bibitem[\protect\citeauthoryear{{Sarazin}}{{Sarazin}}{2002}]{Sarazin2002}
{Sarazin} C.~L.,  2002, in {Feretti} L.,  {Gioia} I.~M.,   {Giovannini} G.,
  eds,  Astrophysics and Space Science Library Vol. 272, \textit{Merging
  Processes in Galaxy Clusters}. pp 1--38 (\mn@eprint {}
  {arXiv:astro-ph/0105418}), \mn@doi{10.1007/0-306-48096-4_1}

\bibitem[\protect\citeauthoryear{{Sarazin}, {Finoguenov}, {Wik}  \&
  {Clarke}}{{Sarazin} et~al.}{2016}]{Sarazin2016}
{Sarazin} C.~L.,  {Finoguenov} A.,  {Wik} D.~R.,   {Clarke} T.~E.,  2016, arXiv
  e-prints, \href {https://ui.adsabs.harvard.edu/abs/2016arXiv160607433S} {p.
  arXiv:1606.07433}

\bibitem[\protect\citeauthoryear{{Sarkar} et~al.,}{{Sarkar}
  et~al.}{2022}]{Sarkar2022}
{Sarkar} A.,  et~al., 2022, \mn@doi [\apjl] {10.3847/2041-8213/ac86d4}, \href
  {https://ui.adsabs.harvard.edu/abs/2022ApJ...935L..23S} {935, L23}

\bibitem[\protect\citeauthoryear{{Simionescu}, {Werner}, {Forman}, {Miller},
  {Takei}, {B{\"o}hringer}, {Churazov}  \& {Nulsen}}{{Simionescu}
  et~al.}{2010}]{Simionescu2010}
{Simionescu} A.,  {Werner} N.,  {Forman} W.~R.,  {Miller} E.~D.,  {Takei} Y.,
  {B{\"o}hringer} H.,  {Churazov} E.,   {Nulsen} P.~E.~J.,  2010, \mn@doi
  [\mnras] {10.1111/j.1365-2966.2010.16450.x}, \href
  {https://ui.adsabs.harvard.edu/abs/2010MNRAS.405...91S} {405, 91}

\bibitem[\protect\citeauthoryear{{Smith} et~al.,}{{Smith}
  et~al.}{2010}]{Smith2010}
{Smith} R.~J.,  et~al., 2010, \mn@doi [\mnras]
  {10.1111/j.1365-2966.2010.17253.x}, \href
  {https://ui.adsabs.harvard.edu/abs/2010MNRAS.408.1417S} {408, 1417}

\bibitem[\protect\citeauthoryear{{Spitzer} \& {Baade}}{{Spitzer} \&
  {Baade}}{1951}]{Spitzer1951}
{Spitzer} Lyman J.,  {Baade} W.,  1951, \mn@doi [\apj] {10.1086/145406}, \href
  {https://ui.adsabs.harvard.edu/abs/1951ApJ...113..413S} {113, 413}

\bibitem[\protect\citeauthoryear{{Stroe}, {Sobral}, {R{\"o}ttgering}  \& {van
  Weeren}}{{Stroe} et~al.}{2014}]{Stroe2014}
{Stroe} A.,  {Sobral} D.,  {R{\"o}ttgering} H. J.~A.,   {van Weeren} R.~J.,
  2014, \mn@doi [\mnras] {10.1093/mnras/stt2286}, \href
  {https://ui.adsabs.harvard.edu/abs/2014MNRAS.438.1377S} {438, 1377}

\bibitem[\protect\citeauthoryear{{Stroe} et~al.,}{{Stroe}
  et~al.}{2015}]{Stroe2015}
{Stroe} A.,  et~al., 2015, \mn@doi [\mnras] {10.1093/mnras/stu2519}, \href
  {https://ui.adsabs.harvard.edu/abs/2015MNRAS.450..646S} {450, 646}

\bibitem[\protect\citeauthoryear{{Taylor}}{{Taylor}}{2005}]{2005ASPC..347...29T}
{Taylor} M.~B.,  2005, in {Shopbell} P.,  {Britton} M.,   {Ebert} R.,  eds,
  Astronomical Society of the Pacific Conference Series Vol. 347, Astronomical
  Data Analysis Software and Systems XIV. p.~29

\bibitem[\protect\citeauthoryear{Taylor \& Babul}{Taylor \&
  Babul}{2004}]{Taylor2004}
Taylor J.~E.,  Babul A.,  2004, \mn@doi [Monthly Notices of the Royal
  Astronomical Society] {10.1111/j.1365-2966.2004.07395.x}, 348, 811

\bibitem[\protect\citeauthoryear{{Tittley} \& {Henriksen}}{{Tittley} \&
  {Henriksen}}{2005}]{Tittley2005}
{Tittley} E.~R.,  {Henriksen} M.,  2005, \mn@doi [\apj] {10.1086/425952}, \href
  {https://ui.adsabs.harvard.edu/abs/2005ApJ...618..227T} {618, 227}

\bibitem[\protect\citeauthoryear{{Urdampilleta}, {Akamatsu}, {Mernier},
  {Kaastra}, {de Plaa}, {Ohashi}, {Ishisaki}  \& {Kawahara}}{{Urdampilleta}
  et~al.}{2018}]{Urdampilleta2018}
{Urdampilleta} I.,  {Akamatsu} H.,  {Mernier} F.,  {Kaastra} J.~S.,  {de Plaa}
  J.,  {Ohashi} T.,  {Ishisaki} Y.,   {Kawahara} H.,  2018, \mn@doi [\aap]
  {10.1051/0004-6361/201732496}, \href
  {https://ui.adsabs.harvard.edu/abs/2018A&A...618A..74U} {618, A74}

\bibitem[\protect\citeauthoryear{{Vaezzadeh} et~al.,}{{Vaezzadeh}
  et~al.}{2022}]{Vaezzadeh2022}
{Vaezzadeh} I.,  et~al., 2022, \mn@doi [\mnras] {10.1093/mnras/stac784}, \href
  {https://ui.adsabs.harvard.edu/abs/2022MNRAS.514..518V} {514, 518}

\bibitem[\protect\citeauthoryear{{Varela} et~al.,}{{Varela}
  et~al.}{2009}]{Varela_2009}
{Varela} J.,  et~al., 2009, \mn@doi [\aap] {10.1051/0004-6361/200809876}, \href
  {https://ui.adsabs.harvard.edu/abs/2009A&A...497..667V} {497, 667}

\bibitem[\protect\citeauthoryear{{Venturi}, {Bardelli}, {Zambelli}, {Morganti}
  \& {Hunstead}}{{Venturi} et~al.}{2001}]{Venturi2001}
{Venturi} T.,  {Bardelli} S.,  {Zambelli} G.,  {Morganti} R.,   {Hunstead}
  R.~W.,  2001, \mn@doi [\mnras] {10.1046/j.1365-8711.2001.04405.x}, \href
  {https://ui.adsabs.harvard.edu/abs/2001MNRAS.324.1131V} {324, 1131}

\bibitem[\protect\citeauthoryear{{Venturi}, {Rossetti}, {Bardelli},
  {Giacintucci}, {Dallacasa}, {Cornacchia}  \& {Kantharia}}{{Venturi}
  et~al.}{2013}]{Venturi2013}
{Venturi} T.,  {Rossetti} M.,  {Bardelli} S.,  {Giacintucci} S.,  {Dallacasa}
  D.,  {Cornacchia} M.,   {Kantharia} N.~G.,  2013, \mn@doi [\aap]
  {10.1051/0004-6361/201322023}, \href
  {https://ui.adsabs.harvard.edu/abs/2013A&A...558A.146V} {558, A146}

\bibitem[\protect\citeauthoryear{{Vijayaraghavan} \& {Ricker}}{{Vijayaraghavan}
  \& {Ricker}}{2013}]{Vijayaraghavan2013}
{Vijayaraghavan} R.,  {Ricker} P.~M.,  2013, \mn@doi [\mnras]
  {10.1093/mnras/stt1485}, \href
  {https://ui.adsabs.harvard.edu/abs/2013MNRAS.435.2713V} {435, 2713}

\bibitem[\protect\citeauthoryear{{Vikhlinin}, {Markevitch}  \&
  {Murray}}{{Vikhlinin} et~al.}{2001}]{Vikhlinin2001}
{Vikhlinin} A.,  {Markevitch} M.,   {Murray} S.~S.,  2001, \mn@doi [\apj]
  {10.1086/320078}, \href
  {https://ui.adsabs.harvard.edu/abs/2001ApJ...551..160V} {551, 160}

\bibitem[\protect\citeauthoryear{{Vulcani} et~al.,}{{Vulcani}
  et~al.}{2018}]{Vulcani2018}
{Vulcani} B.,  et~al., 2018, \mn@doi [\apjl] {10.3847/2041-8213/aae68b}, \href
  {https://ui.adsabs.harvard.edu/abs/2018ApJ...866L..25V} {866, L25}

\bibitem[\protect\citeauthoryear{{Vulcani}, {Poggianti}, {Smith}, {Moretti},
  {Jaff{\'e}}, {Gullieuszik}, {Fritz}  \& {Bellhouse}}{{Vulcani}
  et~al.}{2022}]{Vulcani2022}
{Vulcani} B.,  {Poggianti} B.~M.,  {Smith} R.,  {Moretti} A.,  {Jaff{\'e}}
  Y.~L.,  {Gullieuszik} M.,  {Fritz} J.,   {Bellhouse} C.,  2022, \mn@doi
  [\apj] {10.3847/1538-4357/ac4809}, \href
  {https://ui.adsabs.harvard.edu/abs/2022ApJ...927...91V} {927, 91}

\bibitem[\protect\citeauthoryear{{Walker}, {Fabian}  \& {Kosec}}{{Walker}
  et~al.}{2014}]{Walker2014}
{Walker} S.~A.,  {Fabian} A.~C.,   {Kosec} P.,  2014, \mn@doi [\mnras]
  {10.1093/mnras/stu1996}, \href
  {https://ui.adsabs.harvard.edu/abs/2014MNRAS.445.3444W} {445, 3444}

\bibitem[\protect\citeauthoryear{{Wang}, {Xu}, {Gu}, {Gu}, {Qin}, {Wang},
  {Zhang}  \& {Wu}}{{Wang} et~al.}{2010}]{Wang2010}
{Wang} Y.,  {Xu} H.,  {Gu} L.,  {Gu} J.,  {Qin} Z.,  {Wang} J.,  {Zhang} Z.,
  {Wu} X.-P.,  2010, \mn@doi [\mnras] {10.1111/j.1365-2966.2010.16264.x}, \href
  {https://ui.adsabs.harvard.edu/abs/2010MNRAS.403.1909W} {403, 1909}

\bibitem[\protect\citeauthoryear{{Wang}, {Xu}, {Lee}, {Du}, {Overzier}  \&
  {Shao}}{{Wang} et~al.}{2020}]{Wang2020}
{Wang} J.,  {Xu} W.,  {Lee} B.,  {Du} M.,  {Overzier} R.,   {Shao} L.,  2020,
  \mn@doi [\apj] {10.3847/1538-4357/abb9aa}, \href
  {https://ui.adsabs.harvard.edu/abs/2020ApJ...903..103W} {903, 103}

\bibitem[\protect\citeauthoryear{{Yuan} \& {Han}}{{Yuan} \&
  {Han}}{2020}]{Yuan2020}
{Yuan} Z.~S.,  {Han} J.~L.,  2020, \mn@doi [\mnras] {10.1093/mnras/staa2363},
  \href {https://ui.adsabs.harvard.edu/abs/2020MNRAS.497.5485Y} {497, 5485}

\bibitem[\protect\citeauthoryear{{Yuan}, {Han}  \& {Wen}}{{Yuan}
  et~al.}{2022}]{Yuan2022}
{Yuan} Z.~S.,  {Han} J.~L.,   {Wen} Z.~L.,  2022, \mn@doi [\mnras]
  {10.1093/mnras/stac1037}, \href
  {https://ui.adsabs.harvard.edu/abs/2022MNRAS.513.3013Y} {513, 3013}

\bibitem[\protect\citeauthoryear{{Zabludoff} \& {Zaritsky}}{{Zabludoff} \&
  {Zaritsky}}{1995}]{Zabludoff1995}
{Zabludoff} A.~I.,  {Zaritsky} D.,  1995, \mn@doi [\apjl] {10.1086/309552},
  \href {https://ui.adsabs.harvard.edu/abs/1995ApJ...447L..21Z} {447, L21}

\bibitem[\protect\citeauthoryear{{Zenteno} et~al.,}{{Zenteno}
  et~al.}{2020}]{Zenteno2020}
{Zenteno} A.,  et~al., 2020, \mn@doi [\mnras] {10.1093/mnras/staa1157}, \href
  {https://ui.adsabs.harvard.edu/abs/2020MNRAS.495..705Z} {495, 705}

\bibitem[\protect\citeauthoryear{{Zinger}, {Dekel}, {Birnboim}, {Nagai}, {Lau}
  \& {Kravtsov}}{{Zinger} et~al.}{2018}]{Zinger2018}
{Zinger} E.,  {Dekel} A.,  {Birnboim} Y.,  {Nagai} D.,  {Lau} E.,   {Kravtsov}
  A.~V.,  2018, \mn@doi [\mnras] {10.1093/mnras/sty136}, \href
  {https://ui.adsabs.harvard.edu/abs/2018MNRAS.476...56Z} {476, 56}

\bibitem[\protect\citeauthoryear{{ZuHone}}{{ZuHone}}{2011}]{ZuHone2011}
{ZuHone} J.~A.,  2011, \mn@doi [\apj] {10.1088/0004-637X/728/1/54}, \href
  {http://adsabs.harvard.edu/abs/2011ApJ...728...54Z} {728, 54}

\bibitem[\protect\citeauthoryear{{van Weeren}, {de Gasperin}, {Akamatsu},
  {Br{\"u}ggen}, {Feretti}, {Kang}, {Stroe}  \& {Zandanel}}{{van Weeren}
  et~al.}{2019}]{vanWeeren2019}
{van Weeren} R.~J.,  {de Gasperin} F.,  {Akamatsu} H.,  {Br{\"u}ggen} M.,
  {Feretti} L.,  {Kang} H.,  {Stroe} A.,   {Zandanel} F.,  2019, \mn@doi [\ssr]
  {10.1007/s11214-019-0584-z}, \href
  {https://ui.adsabs.harvard.edu/abs/2019SSRv..215...16V} {215, 16}

\makeatother
\end{thebibliography}

%\begin{thebibliography}{99}
%\end{thebibliography}

%%%%%%%%%%%%%%%%%%%%%%%%%%%%%%%%%%%%%%%%%%%%%%%%%%

%%%%%%%%%%%%%%%%% APPENDICES %%%%%%%%%%%%%%%%%%%%%

\appendix

\section{Centres for computing fractions and visual dynamical state sequence (flowchart).}
\label{sec:centres}
Merging clusters often have several X-ray surface brightness peaks, each possibly associated with a local BCG. Their gas morphology can be highly irregular, and their surface brightness is often low in the X-ray causing the centre determination to be very challenging. In major mergers, the BCG can be significantly displaced from the X-ray peak. 

Also, the determination of BCGs is not always so evident. In most cases, we used \citet{Fasano_2010}'s BCG, which was located near the X-ray centre. The clusters where \citet{Fasano_2010} and this work disagree about the BCGs are only five: Abell 3560, Abell 3158, Abell 1736, Abell 1631a, and Zwicky 8852. After inspecting the images, we found that for A1631a and A3158, the BCG indicated by \citet{Fasano_2010} was not located in the centre of the X-ray emission, whereas the one identified in this work was. The BCGs this work identified in A3560, A3158, A1736, and Z8852 are in the list of cD galaxies, Tab.2, of \citet{Fasano_2010}.
Their coordinates are listed in Tab.~\ref{tab:cls_properties}. In these cases, we adopted this work's BCG as the centre to be consistent with the flowchart in Fig.~\ref{fig:flow_chart}. The BCG identified by \citet{Fasano_2010} and this work for RX1022 is the same (WINGSJ005822.88+265152.6), although it is not the bright galaxy associated with the main X-ray concentration (WINGSJ102209.99+383123.8). We used the latter bright galaxy as a centre to be consistent with our flowchart. A similar case happened in two other clusters. In A2415, we had to replace the BCG (WINGSJ220526.12-054431.1) with the galaxy WINGSJ220538.62-053532.0 located at the centre of the X-ray emission. The other case, A2399, is a very complex merging system identified as a bullet-like cluster by \citet{Lourenco2020}. This work identified more than one central bright galaxy associated with the X-ray emission. We opted to use the galaxy WINGSJ215729.43-074744.1 as the centre.

Below, we list and describe the most challenging clusters for centre determination.

\textbf{Abell 168:} This post-merger system has a very irregular and faint X-ray emission. Previous works identified cold fronts \citep{Hallman2004} and radio relics \citep{Dwarakanath2018}. The radio relics in this cluster are very asymmetrical. After careful inspection of the X-ray and optical images, we chose as the centre the midpoint between the BCG WINGS J011457.58+002551.1, which is associated with the X-ray peak, and the seventh brightest member galaxy (WINGSJ011515.77+001248.6), which seems to be the brightest galaxy associated with the southern X-ray clump centre.   

\textbf{Abell 3376:} This system is one of the most extreme cases of post-merger in our sample. It has two megaparsec-sized arc-shaped radio relics \citep{Bagchi2006}, and its X-ray emission is very asymmetric \citep{Akamatsu2012,Urdampilleta2018}. Simulations \citep{Machado2013} and observations \citep{Durret2013, Monteiro2017b} suggest that this system results from the collision between two clusters with a mass ratio 3–6:1. We used the midpoint between the BCG (WINGSJ060041.09-400240.4) and the bright galaxy (WINGSJ060217.27-395634.2) as the system centre. 

\textbf{Abell 754:} Displaying X-ray shocks and radio relics \citep{Macario_2011}, cold fronts \citep{Ghizzardi2010}, and a halo \citep{Kass1995,Bacchi2003}, this system is yet another case of a post-merger in the plane of the sky. The distribution of galaxies is quite complex \citep{Fabricant1986,Zabludoff1995}. The two main concentrations of galaxies correspond to the two components of diffuse radio emission \citep{Bacchi2003,Inoue2016}. To compute the centre, we used the midpoint between WINGSJ090832.39-093747.3 and WINGSJ090919.21-094158.9 bright elliptical galaxies. 

\textbf{Abell 3667:} Abell 3667 is a bright X-ray source \citep[$5.1 \times 10^{44} \rm erg \ s^{-1}$;][]{Ebeling_1996}, and evidence of merger activity has been observed with various X-ray instruments, e.g., \textit{ROSAT} \citep{Knopp1996}, {\it Chandra} \citep{Vikhlinin2001,Mazzotta2002}, \textit{XMM}-Newton \citep{Briel2004}, Suzaku \citep{Nakazawa2009}, and also in other wavelengths, e.g., radio \citep{Rottgering1997}. The central cluster hosts an X-ray mushroom-shaped feature, probably resulting from the disrupted cool core of a subcluster that passed its centre 1\,Gyr ago \citep{Sarazin2016,Roettiger1999}. Abell 3667 is also one of our sample's most complex interacting systems. It has optical substructures, a cold front, and interesting diffuse radio features such as double relics and an unpolarised bridge associated with a merging event \citep{Rottgering1997}. We used as its centre the computed midpoint between the BCG (WINGSJ201227.32-564936.3) and the galaxy WINGSJ201050.55-564024.5. 

\textbf{Abell 3528:} This double system \citep{Raychaudhury1991} is composed of a northern and southern cluster, A3528-N and A3528-S, respectively, each of them with well-centred BCGs separated by 13.3 arcmin ($\sim 0.45 R_{200}$) in projection and a slight velocity difference ($\sim 357 \,\rm km/s$). Both components are cool core and relaxed \citep{Andrade_Santos_2017,Lopes_2018,Lagana2019}. Optical \citep{Bardelli2001,Baldi2001} and radio observations \citep{Reid1998,Venturi2001} indicate this is a pre-merger system close to virial equilibrium in the central regions of its components. We used the midpoint of the northern (WINGSJ125422.23-290046.8) and southern (WINGSJ125441.01-291339.5) X-ray concentrations BCGs as the system centre.

\textbf{Abell 3716:} Analysing X-ray data from {\it Chandra}, \textit{XMM}-Newton, and \textit{ROSAT} \citep{Andrade2015} concluded that A3716 is a double cluster separated in projection by 400\,kpc. This system shows bi-modality in spatial distribution and position-velocity diagram \citep{Lopes_2018}. \citet{Bilton2019} surmised that the cluster pair is in an early stage of merger by studying its kinematics, which corroborates with a dynamical analysis performed by \citet{Andrade2015} that shows that this system is probably gravitationally bound and will undergo core passage in 500 $\pm$ 200\,Myr. We calculate the system's centre as the midpoint between the BCG of the northern X-ray concentration (WINGSJ205156.94-523746.8) and the southern X-ray concentration (WINGSJ205209.55-524745.4).

\textbf{Abell 3530 and Abell 3532:} This cluster pair is part of the Shapley supercluster and is separated by 22.8 arcmin ($0.95 R_{200}$). The BCG of A3530 is the galaxy WINGSJ125535.99-302051.3, and the BCG of A3532 is WINGSJ125721.97-302149.1. According to \citet{Lakhchaura2013} the two clusters are approaching each other for the first time.

\begin{table*}
\renewcommand{\arraystretch}{1.5}
	\centering
	\caption{\citetalias{Poggianti2016} clusters that cover $0.7\,R_{200}$}
	\label{tab:cls_properties}
	\begin{tabular}{lcccccccccccc} % eight columns, alignment for each
	\hline
Cluster & {\it Chandra} & {\it XMM-Newton} & Spec. complete & SCF & MCF & Relic & Halo & Flowchart &
RA$_{fc}$ & DEC$_{fc}$ & $F_{RPS}^{phot}$ & $F_{RPS}^{spec}$ \\
 & & & & & & & & & (J2000)&(J2000) & & \\ 

\hline
  A85 & yes & yes & yes & c & a &  &  & 3 & 10.46 & -9.30 & 0.06 & 0.07\\
  A133 & yes & yes & yes & l &  &  &  & 3 & 15.67 & -21.88 & 0.31 & 0.57\\
  A147 & no & no & no &  &  &  &  &  & 17.05 & 2.19 & 0.07 &\\
  A151 & no & yes & yes &  &  &  &  & 2 & 17.21 & -15.41 & 0.06 & 0.13\\
  A160 & yes & yes & yes &  &  &  &  & 3 & 18.25 & 15.49 & 0.05 & 0.06\\
  A168 & no & yes & yes &  & e & f &  & 5 & 18.78 & 0.32 & 0.12 & 0.16\\
  A193 & yes & no & yes &  &  &  &  & 2 & 21.28 & 8.70 & 0.00 & 0.00\\
  A500 & no & yes & yes &  &  &  &  & 3 & 69.72 & -22.11 & 0.22 & 0.19\\
  A602 & no & yes & yes &  &  &  &  & 4 & 118.36 & 29.36 & 0.00 & 0.00\\
  A754 & yes & yes & yes &  & a & f & f & 5 & 137.23 & -9.66 & 0.00 & 0.00\\
  A957x & yes & no & yes &  &  &  &  & 3 & 153.41 & -0.93 & 0.26 & 0.21\\
  A1069 & no & yes & yes &  &  &  &  & 4 & 159.93 & -8.69 & 0.00 & 0.00\\
  A1291 & no & no & yes &  &  &  &  &  & 173.10 & 55.97 & 0.00 & 0.00\\
  A1631a & no & no & yes &  &  &  &  &  & 193.33 & -15.53 & 0.21 & 0.14\\
  A1668 & yes & no & yes &  &  &  &  & 3 & 195.94 & 19.27 & 0.37 & 0.29\\
  A1795 & yes & yes & yes & k &  &  &  & 3 & 207.22 & 26.59 & 0.13 & 0.14\\
  A1831 & yes & yes & yes &  &  &  &  & 3 & 209.81 & 27.98 & 0.19 & 0.00\\
  A1983 & no & yes & yes &  &  &  &  & 2 & 223.23 & 16.70 & 0.13 & 0.12\\
  A1991 & yes & yes & yes &  &  &  &  & 2 & 223.63 & 18.64 & 0.20 & 0.19\\
  A2107 & yes & no & yes &  &  &  &  & 2 & 234.91 & 21.78 & 0.00 & 0.00\\
  A2124 & yes & yes & yes &  &  &  &  & 2 & 236.25 & 36.11 & 0.27 & 0.43\\
  A2149 & no & no & yes &  &  &  &  &  & 240.37 & 53.95 & 0.25 & 0.86\\
  A2169 & no & no & yes &  &  &  &  &  & 243.49 & 49.19 & 0.00 & 0.00\\
  A2382 & no & no & yes &  &  &  &  &  & 327.98 & -15.71 & 0.04 & 0.00\\
  A2399 & no & yes & yes &  & g &  &  & 4 & 329.37 & -7.80 & 0.16 & 0.14\\
  A2415 & yes & yes & yes &  &  &  &  & 4 & 331.41 & -5.59 & 0.48 & 0.34\\  
  A2457 & yes & no & yes &  &  &  &  & 3 & 338.92 & 1.48 & 0.00 & 0.00\\
  A2593 & yes & no & yes &  &  &  &  &   & 351.08 & 14.65 & 0.10 & 0.00\\
  A2626 & yes & yes & yes & i &  &  &  & 3 & 354.13 & 21.15 & 0.12 & 0.19\\
  A2657 & yes & yes & yes &  &  &  &  & 3 & 356.24 & 9.19 & 0.00 & 0.00\\
  A2717 & yes & yes & yes &  &  &  &  & 2 & 0.80 & -35.94 & 0.00 & 0.00\\
  A2734 & yes & yes & yes &  &  &  &  & 2 & 2.84 & -28.85 & 0.84 & 0.00\\
  A3128 & yes & yes & yes &  &  &  &  & 3 & 52.46 & -52.58 & 0.00 & 0.00\\  
  A3158 & yes & yes & yes &  & m &  &  & 4 & 55.72 & -53.63 & 0.28 & 0.14\\
  A3266 & yes & yes & yes &  & a & n & n & 5 & 67.81 & -61.45 & 0.04 & 0.04\\
  A3376 & yes & yes & yes &  & j & f &  & 5 & 90.37 & -39.99 & 0.08 & 0.08\\
  A3490 & no & yes & no &  &  &  &  & 3 & 176.34 & -34.43 & 0.47 & \\
  A3528 & yes & yes & yes &  &  &  &  & 1 & 193.63 & -29.12 & 0.00 & 0.00\\
  A3530/32 & yes & yes & yes &  &  &  &  & 1 & 194.12 & -30.36 & 0.18 & 0.17\\
  A3556 & no & no & yes &  &  &  &  &   & 201.03 & -31.67 & 0.00 & 0.00\\

    \end{tabular}
    \end{table*}
    \begin{table*}
    \renewcommand{\arraystretch}{1.5}
	\centering
    \begin{tabular}{lcccccccccccc}
Cluster & {\it Chandra} & {\it XMM-Newton} & Spec. complete & SCF & MCF & Relic & Halo & Flowchart &
RA$_{fc}$ & DEC$_{fc}$ & $F_{RPS}^{phot}$ & $F_{RPS}^{spec}$ \\
 & & & & & & & & & (J2000)&(J2000) & & \\ 
 \hline
  A3558 & yes & yes & yes & b &  &  &  & 3 & 201.99 & -31.50 & 0.05 & 0.07\\
  A3560 & yes & yes & yes & d &  &  &  & 3 & 203.18 & -33.14 & 0.10 & 0.10\\
  A3667 & yes & yes & yes &  & a & f &  & 5 & 302.91 & -56.75 & 0.06 & 0.07\\
  A3716 & yes & yes & yes &  &  &  &  & 1 & 313.01 & -52.71 & 0.10 & 0.00\\
  A3809 & yes & no & yes &  &  &  &  & 3 & 326.75 & -43.90 & 0.00 & 0.00\\
  A3880 & yes & no & yes &  &  &  &  & 2 & 336.98 & -30.58 & 0.11 & 0.14\\
  A4059 & yes & yes & yes & h &  &  &  & 3 & 359.25 & -34.76 & 0.06 & 0.09\\
  RX1022 & yes & yes & no &  &  &  &  & 3 & 155.54 & 38.52 & 0.00 &\\
  RX1740 & no & yes & yes &  &  &  &  & 2 & 265.13 & 35.65 & 0.00 & \\
  MKW3s & yes & yes & yes &  &  &  &  & 2 & 230.47 & 7.71 & 0.00 & 0.00\\
  Z2844 & no & yes & yes &  &  &  &  & 2 & 150.65 & 32.71 & 0.00 & 0.00\\
  IIZW108 & yes & no & yes &  &  &  &  & 3 & 318.48 & 2.57 & 0.00 & 0.00\\  
  \hline
  \multicolumn{3}{l}{$^a$ \citet{Ghizzardi2010}}\\
  \multicolumn{3}{l}{$^b$ \citet{Rossetti2007}}\\
  \multicolumn{3}{l}{$^c$ \citet{Ichinohe2015}}\\ 
  \multicolumn{3}{l}{$^d$ \citet{Venturi2013}}\\ 
  \multicolumn{3}{l}{$^e$ \citet{Hallman2004}}\\  
  \multicolumn{3}{l}{$^f$ \citet[][ and references therein]{vanWeeren2019}}\\  
  \multicolumn{3}{l}{$^g$ \citet{Mitsuishi2018}}\\   
  \multicolumn{3}{l}{$^h$ \citet{Lagana2010}}\\  
  \multicolumn{3}{l}{$^i$ \citet{Ignesti2018}}\\   
  \multicolumn{3}{l}{$^j$ \citet{Urdampilleta2018}}\\  
  \multicolumn{3}{l}{$^k$ \citet{Walker2014}}\\    
  \multicolumn{3}{l}{$^l$ \citet{Randall2010}}\\   
  \multicolumn{3}{l}{$^m$ \citet{Wang2010}}\\
  \multicolumn{3}{l}{$^n$ \citet{Riseley2022}}\\  
%\hline
\end{tabular}
\end{table*}
%If you want to present additional material which would interrupt the flow of the main paper, it can be placed in an Appendix which appears after the list of references.

%%%%%%%%%%%%%%%%%%%%%%%%%%%%%%%%%%%%%%%%%%%%%%%%%%

% Don't change these lines
\bsp	% typesetting comment
\label{lastpage}
\end{document}